\documentclass[11pt]{article}
\usepackage{amsfonts}
\usepackage{amssymb}
\usepackage{mathrsfs}
\usepackage{graphicx}
\usepackage{amsmath,amsthm,amssymb,amscd}
\usepackage[all]{xy}
\usepackage[notref,notcite]{}
\usepackage{subfig}
\usepackage{hyperref}
\usepackage{multirow}

\usepackage[dvips]{color}
\usepackage{amsfonts}
\usepackage{float}
\def\eqa{\begin{eqnarray}}
\def\eqae{\end{eqnarray}}
\def\eq{\begin{equation}}
\def\eqe{\end{equation}}
\def\be{\begin{equation}}
\def\ee{\end{equation}}
\def\bea{\begin{eqnarray}}
\def\eea{\end{eqnarray}}
\def\ba{\begin{array}}
\def\ea{\end{array}}
\def\bd{\begin{displaymath}}
\def\ed{\end{displaymath}}

\def\>{\rangle}
\def\<{\langle}

\def\e{\epsilon}           

\def\NPB{Nucl.\ Phys.\ B}

\def\ZPC{Z.\ Phys.\ C}

\newcommand{\beqas}{\begin{eqnarray*}}
\newcommand{\beqa}{\begin{eqnarray}}

\newcommand{\eeqas}{\end{eqnarray*}}
\newcommand{\eeqa}{\end{eqnarray}}

\newcommand{\ga}{\gamma}

\newcommand{\ka}{\kappa}
\newcommand{\la}{\lambda}

\newcommand{\De}{\Delta}

\numberwithin{equation}{section}

\begin{document}

\begin{titlepage}
\begin{flushright}
PUPT 2459 \\
\end{flushright}
\vspace{1cm}

\begin{center}
{\Large \bf Anomalies without an action}\\[1cm]
Wei-Ming Chen$^{a}$, Yu-tin Huang$^{a,b}$, and David A. McGady$^c$ 
~\\[5mm]

$^a${\it Department of Physics and Astronomy}\\
 {\it National Taiwan University, Taipei}\\
 {\it Taipei 10617, Taiwan, R.O.C}\\
~\\
$^b${\it School of Natural Sciences}\\
 {\it Institute for Advance Study}\\
 {\it Princeton, NJ 08540, USA}\\
~\\
$^c${\it Department of Physics, Jadwin Hall}\\
{\it Princeton University}\\
{\it Princeton, NJ 08544, USA}\\
~\\

~\\
~\\
\today
~\\
~\\

\end{center}

\abstract{Modern on-shell methods allow us to construct both the classical and quantum S-matrix for a large class of theories, without utilizing knowledge of the interacting Lagrangian. It was recently shown that the same applies for chiral gauge theories, where the constraints from anomaly cancelation can be recast into the tension between unitarity and locality, without any reference to gauge symmetry. In this paper, we give a more detailed exploration, for chiral QED and QCD. We study the rational terms that are mandated by locality, and show that the factorization poles of such terms reveal a new particle in the spectrum, the Green-Schwarz two-from. We further extend the analysis to six-dimensional gravity coupled to chiral matter, including self-dual two-forms for which covariant actions generically do not exist. Despite this, the on-shell methods define the correct quantum S-matrix by demonstrating that locality of the one-loop amplitude requires combination of chiral matter that is consistent with that of anomaly cancelation.
}

\end{titlepage}

\tableofcontents
\newpage

\section{Introduction and summary of results} 
\label{introduction}
Analyticity has long been known to impose stringent constraints on the S-matrix, and, from this, non-trivial predictions can be drawn for the underlying effective field theory. Classic examples include Coleman and Grossman's proof of 't Hoofts anomaly matching conditions~\cite{Coleman:1982yg}, as well as Adams' et al.~\cite{ Adams:2006sv} proof of positivity for signs of the leading irrelevant operator. For the past decade, it has been understood that for a large class of gauge and gravity theories, systematic use of such constraints in the form of generalized unitarity methods~\cite{UnitarityMethod} and recursion relations~\cite{BCFW}, one retains sufficient information to fully determine the amplitude iteratively from the fundamental building block: the three-point amplitude. For massless theories, with a given mass-dimension of the coupling constant the latter is in fact unique. 

The complete irrelevance of the interaction Lagrangian in these developments has an appealing feature for the following reason: there are a plethora of interesting interacting theories for which no action is known. That this is so has been attributed to various reasons such as the absence of small expansion parameter, the inability to non-abelianize the underlying gauge symmetry of the asymptotic states, et cetera. However, it serves to remind ourselves that not too long ago such statements were widely used for the world-volume theory of multiple M2 branes, prior to the ascent of BLG~\cite{BLG} and ABJM~\cite{ABJM} theories. Thus the removal of the ``off-shell" yoke is welcoming. Now instead of asking if one can write down an interacting Lagrangian, we ask if one can construct a consistent S-matrix for the asymptotic states of the theory. Any difficulty encountered in the process will be independent of any off-shell formulation. Indeed it has been shown that for both BLG and ABJM theories, global symmetries as well as factorization constraints is sufficient to completely determine the tree-level amplitude~\cite{ABJMD, BLGYT}. 

However, gauge and gravitational anomalies also impose constraints on the quantum consistency of chirally coupled gauge and gravitational theories, and one might doubt whether the on-shell approach, being blind to any gauge symmetry, fails to capture such inconsistencies. The on-shell manifestation of such inconsistency is usually stated as the non-decoupling of the longitudinal degrees of freedom. Unfortunately such discussion utilizes formal polarization (tensors) vectors and Feynman rules, for which an action is a prerequisite. Such discussion also appears to be incompatible with the modern on-shell program, where only the physical degrees of freedom are present throughout the construction. At this point, one might be tempted to conclude that unitarity methods are ill-fitted with chiral theories, and a completely on-shell construction is simply too much to ask for these theories. However, this cannot be the case, since if one were to blindly construct the quantum S-matrix using generalized unitarity methods, in absence of any sickness, one could simply declare that a consistent S-matrix is obtained, as the construction manifestly respects unitarity. One would then move on to conclude that there is no perturbative inconsistency for chiral gauge theories. Thus there must be a more physical manifestation of such sickness that does not rely on any notion of gauge symmetry or unphysical degrees of freedom.    

Generalized unitarity utilizes the fact that the discontinuity of branch cuts are given by the product of lower loop-order amplitudes. Using sufficiently many cuts, one can fully determine the integrand for the cut constructible part of the amplitude. Since by construction, this method correctly reproduces all branch cuts, the amplitude manifestly respects unitarity. However, this does not guarantee locality. Here locality is defined by the property that the only singularity in the amplitude is associated with propagator-like singularities, $1/(p_i+p_j+\cdots+p_l)^2$. The unitarity methods determine the amplitude up to rational terms which are free of branch cuts.\footnote{Although one can always embed the theory in higher dimensions and use the higher-dimensional unitarity cuts to fix such rational terms~\cite{DGenCuts}, this cannot be done for chiral theories which have no higher-dimensional parent.} However, such terms may have singularities, and its role is precisely to ensure that the full amplitude respects locality~\cite{BernReview}. 

Not surprisingly, recently two of the authors~\cite{YTDM} revealed that for chiral-gauge theories, the result obtained from unitarity methods violates locality, and is unrepairable unless certain conditions are met. More precisely, by enforcing locality on the cut-constructible result, the requisite rational term inevitably introduces new factorization channels. In four-dimensional Yang-Mills theory coupled to chiral fermions, the residue of this factorization channel cannot be interpreted as a product of tree-amplitudes: the factorization channel is therefore unphysical. Requiring the absence of such factorization-channels exactly leads to the well known anomaly cancellation condition, $d^{abc}=0$! In six-dimensions, the same applies, except that now the residue of the new factorization channel can be interpreted as the product of two three-point amplitudes, each corresponding to a two-from coupling to two gauge fields. Thus one recovers the full anomaly cancellation conditions in six-dimensions: the requirement of locality requires either $str(T_1T_2T_3T_4)=0$, or that it factorizes. For the latter, the factorization channel automatically introduces a new state to the spectrum, the two-form of the Green-Schwarz mechanism~\cite{GS, GS6D}!  Thus, in short, inconsistency of chiral-theories arises as the incompatibility of unitarity constraints and that of locality.

In this paper, we expand on the analysis for six-dimensional chiral QCD, and extend it to chiral QED and gravity. For the gauge theories, while locality mandates a unique rational term which introduces a new factorization channel, the exchanged particle is identified to be a two-form only via dimensional and little-group analysis. Here we explicitly show that by gluing three-point amplitudes of a two-form coupled to two gauge-fields, one reproduces the exact residue that appeared in the factorization channel of the complete chiral-fermion loop-amplitude. This completes the proof that on-shell methods automatically includes the requisite spectrum for a local and unitary quantum field theory, even if the initial spectrum is incomplete. Conversely, we also demonstrate that standing on its own, parity-odd rational terms cannot have consistent factorization in all channels, if any one of them involve the exchange of a two-form between vector fields. This is an on-shell manifestation of the well known fact that if the two-form couples to the gauge fields via $H^2=(dB+AdA)^2$ and $B\wedge F\wedge F$, one cannot simultaneously maintain gauge invariance for both operators. 

Remarkably, direct Feynman diagram calculation reveals that the rational term obtained from on-shell methods exactly matches that coming from the combination of the anomalous rational term from the loop-amplitude and that from the Green-Schwarz mechanism. It is interesting that whilst the anomaly is completely canceled, a remnant remains in the form of a parity-odd gauge-invariant rational term, whose presence is perhaps an afterthought in the traditional point of view. However, from the amplitude point of view, such rational terms are of upmost importance as they are mandated by locality, whilst anomaly cancellation is merely an ``off-shell" mechanism to reproduce the requisite rational terms. As a side result, we also present the anomalous rational terms for chiral gauge theories in arbitrary even dimensions.

Chirally coupled gravity presents an interesting test case for our approach. In particular, in six-dimensions, no covariant action exists which couples (a generic number of) self-dual two-forms to gravity. In fact, in the original work of Alvarez-Gaume and Witten~\cite{AlvarezGaume:1983ig}, this has been presented as an argument for the possibility of a gravitational anomaly. Here we demonstrate that generalized unitarity methods do define a consistent quantum S-matrix. To begin, we explicitly construct the parity-odd amplitude for the four-point one-loop amplitude, associated with chiral-matter. Consistency is demonstrated by showing that the conditions imposed by locality of the quantum S-matrix precisely match the gravitational anomaly cancellation conditions. Once these conditions are satisfied, we obtain the unitary and local gravity amplitude associated with chiral matter. 

This paper is organized as follows. In section 2, we review generalized unitarity, with an emphasis on the role of rational terms in enforcing locality. In section 3, we show that parity-odd loop-amplitudes constructed from generalized unitarity, generically requires rational terms that introduces new factorization channels, and thus potential violation of locality. Enforcing locality, in this setting, requires the same group theory constraints as the traditional anomaly cancellation. In section 4, we explicitly show, that the ``new'' poles which appear in \emph{local} parity-odd loop amplitude in chiral gauge-theories must be identified with an exchange of a two-form with parity-violating couplings to gauge-fields. In section 5, we revisit anomaly cancellation from the point of view of Feynman diagrams, and demonstrate that the rational term obtained from the unitarity methods are in fact a combination of the anomalous rational terms from one-loop chiral fermion amplitude and the tree-level Green-Schwarz mechanism. Finally, in section 6, we make contact with the existence of, and recover cancellation conditions for, perturbative gravitational anomalies in six-dimensions.

\section{Cut constructibility and the rational terms }
Generalized unitarity~\cite{UnitarityMethod}, by construction, furnishes a representation for the loop-integrand which reproduces the correct unitarity-cuts. Within the context of an integral basis, the perturbative structure of a particular S-matrix element is encoded in the coefficients of each integral. Depending on the nature of the theory, different integral bases manifest different properties. At one-loop, for general purposes it is convenient to use a basis of scalar integrals. Here, one-loop amplitudes in a generic $D$-dimensional massless theory---in this paper $D$ is always taken to be integer---decompose into:
\eq
A^{\rm 1-Loop}=\sum_{i_D} c^{i_D}_D I^i_D+ \sum_{i_{D-1} } c_{D-1}^{i_{D-1}} I_{D-1}+\cdots + \sum_{i_{D-1} } c_2^{i_{2}} I_{2}+R+\mathcal{O}(\epsilon) \, ,
\label{IntegralBasis}
\eqe
where $I^{i_a}_a$ represent scalar integrals with $a$ propagators, diagrammatically an $a$-gon, and $i_a$ labels the different distinct $a$-gons. The coefficient in front of each integral is to be fixed by unitarity cuts. For a review of why such an integral basis is sufficient for all one-loop amplitudes, we refer the reader to~\cite{LanceReview} for a brief discussion. The last term $R$ indicates a possible rational function that is cut-free.

An important subtlety is whether the unitarity cuts are defined in $D$ or $D-2\epsilon$-dimensions. Since the scalar loop-integrals in eq.\eqref{IntegralBasis} do not contain any loop momentum dependence in the numerator, their coefficients are blind to whether or not the unitarity cut is defined in $D$ or $D-2\epsilon$-dimensions. On the other hand, the rational terms $R$, can be obtained via  $D-2\epsilon$-dimensional unitarity cuts~\cite{DGenCuts}, which requires tree-amplitudes within the unitarity-cut to be analytically continued into higher dimensions. While this approach for obtaining the rational terms is widely applied in QCD, it becomes problematic for chiral theories or theories with Chern-Simons terms. Finally, to properly extend the theory to higher dimensions, one requires the knowledge of the explicit action, which is antonymous to the program of the on-shell S-matrix.

It is then useful to take a step back and ask: what physical principle mandates the existence of these rational terms? Specifically, given that the result in eq.\eqref{IntegralBasis}, absent the rational terms $R$, is manifestly unitary in all channels, in what sense are the cut-constructed terms without their rational counterparts incorrect? To answer this question, it is useful to consider an explicit example. Let us consider a fundamental fermion contribution to the four-point one-loop gluon amplitude in four-dimensions. Applying standard unitary cut methods~\cite{FordeRules}, the integral coefficients for $A^{\rm 1-Loop}(1^+2^-3^+4^-)$ from this fermion loop:
\eqa
\nonumber \vcenter{\hbox{\includegraphics[scale=0.45]{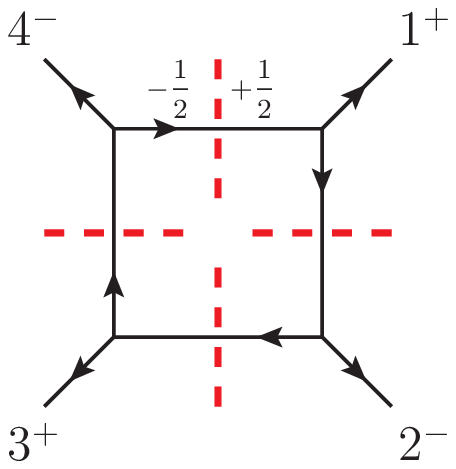}}}\;\;c_4=-\frac{t^4s^2}{u^4} &&\vcenter{\hbox{\includegraphics[scale=0.45]{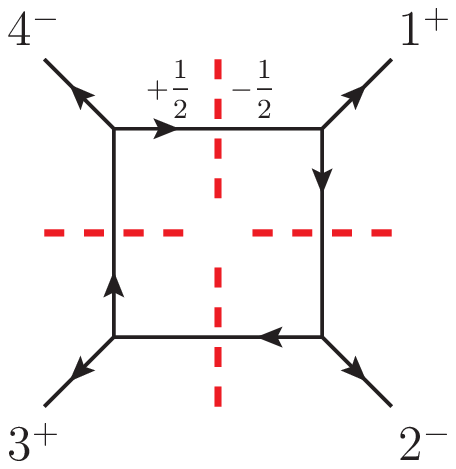}}}\;\;c_4=-\frac{s^4t^2}{u^4} \\ 
\vcenter{\hbox{\includegraphics[scale=0.45]{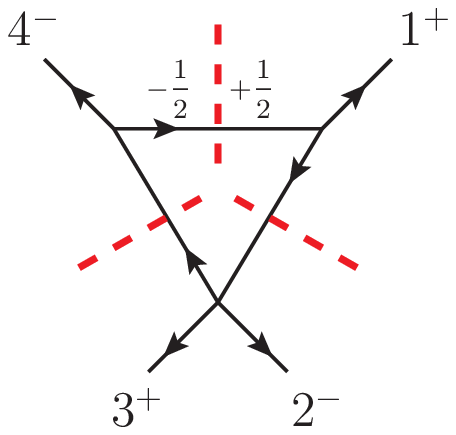}}}\;\;c^{(23)}_3= \frac{t^4s}{u^4}&&\vcenter{\hbox{\includegraphics[scale=0.45]{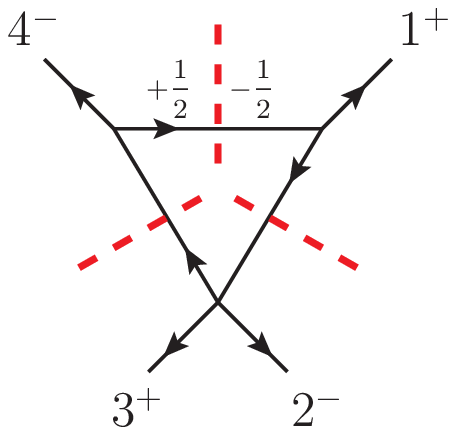}}}\;\;c^{(23)}_3= \frac{t^2s^3}{u^4}\quad\;\,\\ 
\nonumber \vcenter{\hbox{\includegraphics[scale=0.45]{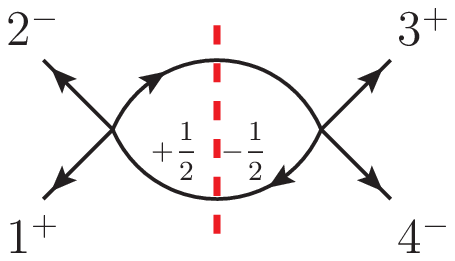}}}\;\;c^{(12)}_2=\frac{t(2 t^2-s^2 - 5 s t )}{6u^3}&&\vcenter{\hbox{\includegraphics[scale=0.45]{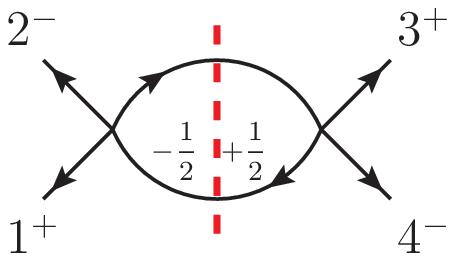}}}\;\;c^{(12)}_2=\frac{t( 2 t^2+11 s^2 + 7 s t )}{6u^3}
\eqae
where the superscript in $c^{(ij)}$ denotes the legs on the massive corners. Note: for future convenience, the contributions to the integral coefficients from each of the two different fermion helicities circling in the loop have been separated. This separation facilitates finding the even- and odd-parity components of the loop-amplitude. The other orderings can be obtained by cyclic rotation. Combined with the integrated scalar integrals, we find the following fermion contribution to the $A^{\rm 1-L}(1^+2^-3^+4^-)$ amplitude : 
\eqa
\nonumber A_{1/2,\,Cut}^{1-{\rm L, even}}(1^+2^-3^+4^-)&=&A^{\rm tree}(1^+2^-3^+4^-)\bigg[\frac{-(-s)^{-\e}-(-t)^{-\e}}{3\epsilon}-\frac{st(s^2+t^2)}{2u^4}(\log^2x+\pi^2) \\
&&-\frac{(s-t)(s^2-st+t^2)}{3u^3}\log x\bigg]\,,
\label{even}
\eqae
where $x\equiv t/s$, the subscript in $A_{1/2}$ indicates it is a fermion loop contribution, $Cut$ indicates it is the result derived from purely four-dimensional unitarity cuts, and ``${\rm even}$'' indicates we have taken the parity-even combination of the various unitarity-cuts.\footnote{Notice that the $-2/3$ in front of the $1/\epsilon$ UV-divergence is precisely the minus of the one-loop beta function, as expected. See~\cite{Peng} and references therein.}

We would now like to ask, what precisely is wrong with eq.(\ref{even})? The branch-cut structure is guaranteed to be correct by construction, and thus unitarity is not violated. However, the amplitude not only has to satisfy unitarity, but locality as well. Locality requires that the loop amplitude can only have propagator singularities, and that the residue on these propagator-poles must be given by the product of lower-order amplitudes (lower-order means reduced number of legs and reduced loop-order):
\eq\label{FactorRule}
A_n^{\ell}|_{K^2\rightarrow 0} =\sum_{\ell_1,\ell_2} A^{\ell_1}_{n+1-k}\frac{1}{K^2}A^{\ell_2}_{k+1}+\mathcal{O}((K^2)^0)\,,
\eqe
where $\ell$, $\ell_1$, and $\ell_2$ denote loop-order and are subject to the constraint $\ell = \ell_1+\ell_2$. 

Going back to eq.(\ref{even}) we see that $1/u^n$ factors are ubiquitous in each term, and thus provide potential violations of locality. Expanding eq.(\ref{even}) around $u=0$ one finds:
\eq
 A_{1/2,\,Cut}^{1-{\rm L, even}}(1^+2^-3^+4^-)\bigg|_{u\rightarrow0}=A^{\rm tree}(1^+2^-3^+4^-)\left(-\frac{s^2}{u^2}-\frac{s}{u}+\mathcal{O}(u^0)\,\right)\,.
\eqe
Since the spurious poles have rational residues, the result in eq.(\ref{even}) violates locality! Given that the part of the amplitude containing cuts is completely determined, the only way this may be remedied is through addition of a cut-free term to the amplitude to cancel this singularity: a rational term. Dimensional analysis and cyclic symmetry entirely fix the rational term needed to correct this $1/u^2$-singularity to be,
\eq
R^{\rm even}(1^+2^-3^+4^-)=-A^{\rm tree}(1^+2^-3^+4^-)\frac{st}{u^2} \,,
\label{nonCR}
\eqe
Note that, since this is a color-ordered amplitude, a $1/u$ pole with is unphysical. Luckily $R^{\rm even}$ automatically removes the $1/u$ pole as well. Thus the unique unitary and local answer is: 
\eq
A_{1/2}^{1-{\rm L, even}}= A_{1/2,\,Cut}^{\rm 1-Loop, even}+R^{\rm even}\,,
\eqe
reproducing the correct answer obtained, long ago, in \cite{BernQCD}\,. 

From the above discussion we see that the role of rational terms is to correct any non-locality introduced into the results from $D$-dimensional unitarity cuts. Note that this implies that the rational terms must have certain factorization properties, which allow it to be obtained through recursion relations, as demonstrated in ~\cite{RationalRecur1, RationalRecur2, RationalRecur3}. One might wonder if it is possible to add a term that is simply a constant times the tree-amplitude. This is not an acceptable modification to the loop amplitude: the tree-amplitude contains $s$- and $t$-factorization channels, and the presence of such a term would lead to a one-loop modification to the three-point on shell amplitudes, which are not supposed to be corrected in perturbation theory. Note that $R^{\rm even}$ avoids this problem as its numerator is proportional to $st$, which vanishes on either of these two factorization channels. This subtlety will rear its head when we consider chiral fermions, and is the avatar of four-dimensional anomalies as we discuss in the next section.

\section{Anomalies as spurious poles}
In the section, we will repeat the process discussed above for chiral gauge theories. That is, we will construct parity-odd contributions to unitarity cuts of gauge-theory amplitudes due to chiral fermions, in four- and six-dimensions. 
We will study the pole structure of the unitarity cut result, and determine whether rational terms are needed to insure locality. As we will see, there is a subtlety in the  prerequisite rational terms that plants the seed for the inconsistency of chirally coupled gauge and, in a later section, gravity theories coupled to chiral matter.  

\subsection{D=4 QCD\label{sec:4DChiralQCD}}
Let us now revisit the same scattering process fermion-loop, and now focus on the parity-odd configuration, where we take the \textit{the difference} of the two helicity configurations in each unitarity cut. Note that this part of the amplitude is only present in chiral theories. The result is:
\eqa
\nonumber  && A_{1/2,\,Cut}^{1-{\rm L, odd}}(1^+2^-3^+4^-)=A^{\rm tree}(1^+2^-3^+4^-)\left[-\frac{st(s-t)}{2u^3}(\log^2x+\pi^2)-\left(\frac{2st}{u^2}\right)\log x\right] \,.
\label{odd}
\eqae
Note that the above result is only cyclic invariant up to a sign. This is because we are using helicity basis in combination with the Levi-Cevita tensor. Indeed one can identify:
\eqa
\nonumber A^{\rm tree}(1^+2^-3^+4^-)(s-t)&=&-2u\left[(F_1\wedge F_3)\left(\frac{\epsilon_4\cdot k_2}{u}-\frac{\epsilon_4\cdot k_3}{s}\right)\left(\frac{\epsilon_2\cdot k_4}{u}-\frac{\epsilon_2\cdot k_1}{s}\right)\right.\\
\nonumber&+&\left.(F_2\wedge F_4)\left(\frac{\epsilon_1\cdot k_3}{u}-\frac{\epsilon_1\cdot k_4}{t}\right)\left(\frac{\epsilon_3\cdot k_1}{u}-\frac{\epsilon_3\cdot k_2}{t}\right)\right]+cyclic\,,
\label{Tensor}
\eqae 
which reveals the cyclic structure outside of a particular helicity assignment. To see if rational terms are needed, we look at the residue of the apparent spurious $u$-pole:
\eq\label{oddU}
A_{1/2,\,Cut}^{1-{\rm L, odd}}(1^+2^-3^+4^-)|_{u\rightarrow0}=A^{\rm tree}(1^+2^-3^+4^-)\left(-\frac{s}{u}+\mathcal{O}(u^0)\right)\,.
\eqe 
Locality again requires the absence of such spurious poles through addition of a rational term. Taking into account the fact that the amplitude attains a minus sign under cyclic shift, the requisite parity-odd rational term is:
\eq
R^{\rm \,odd}(1^+2^-3^+4^-)=A^{\rm tree}(1^+2^-3^+4^-)\frac{s-t}{2u} = \langle24\rangle^2 [13]^2\frac{s-t}{2stu} \,.
\label{CR}
\eqe

Naively the following sum is fully unitary and local: 
\eq\label{TempAnsw}
A_{1/2,\,Cut}^{1-{\rm L, odd}}+R^{\rm \,odd}\,.
\eqe
However, as is plain from Eq.(\ref{CR}), this new parity-odd rational term introduces non-trivial contributions to the $s$- and $t$-channel residues. This contrasts sharply with the $R^{\rm \,even}$. Herein lay the seeds of inconsistencies in parity-violating gauge theories: the rational terms, required for locality in the parity-odd amplitude, introduce \emph{new} corrections to residues on the $s$- and $t$- poles. Since the dimension of the residue is $2$, for eq.(\ref{FactorRule}) to be respected the residue must be given by either (a) the product of two mass-dimension 1 three-point amplitudes, or (b) a product of mass-dimension zero and a mass-dimension two tree amplitude. Let us consider each case in detail:
\begin{itemize}
  \item (1,1): We consider possible mass-dimension one three-point amplitudes with two vectors of opposite helicity and one unknown particle specie $A_3(g^+g^- *)$. A simple analysis of helicity constraint tells us that the unknown particle can only be a vector. Thus the only three-point amplitude allowed is the MHV, or $\overline{\rm MHV}$, amplitude which does not have one-loop corrections.  
  \item (2,0): Note that there is a mass-dimension two amplitude involving two gluons and one scalar, generated by the operator $\phi F^2$. However there are no mass-dimension zero amplitudes involving two gluons.\cite{4pointMassless}
\end{itemize}
Thus from the above analysis we conclude that eq.(\ref{TempAnsw}) does not respect locality, as incarnated in eq.(\ref{FactorRule}): theories which chirally couple fermions to gauge bosons have an inherent tension between locality and unitarity!

However, there is a potential way out. Since the fully color-dressed amplitude is given by a sum of single-trace amplitudes (we will assume fundamental fermions for now), it is possible that the new factorization poles introduced by the rational terms actually cancel in the fully dressed amplitude. There are six distinct single trace structures, and the rational term in the fully color-dressed amplitude is given by:
\eqa
\nonumber \mathcal{R}&=&\frac{\langle24\rangle^2 [13]^2}{2stu}[(s-t)Tr(1234)+(u-s)Tr(1342)+(t-u)Tr(1423)\\
&&~~~~~~~~~~~~+(s-u)Tr(1243)+(u-t)Tr(1324)+(t-s)Tr(1432) ] \,.
\eqae
Let us consider the residue for the $s\rightarrow0$. The contribution from $Tr(1234)$, $Tr(1342)$, $Tr(1432)$ and $Tr(1243)$ sum to:
\eq\label{4DTrace}
\frac{\langle24\rangle^2 [13]^2}{2su}[-Tr(1234)-Tr(1342)+Tr(1243)+Tr(1432)]\,.
\eqe
Note that we did not include the contribution from $Tr(1423)$ and $Tr(1324)$, since the $s$-channel pole in these two trace structures by construction cancels against the $s$-pole from the cut-constructed part of the amplitude. Thus if the group theory structure in eq.(\ref{4DTrace}) vanishes, one obtains a fully consistent amplitude. This constraint on the group theory factor can be rewritten as:
\eqa
\hspace{-0.2cm}Tr(1432)-Tr(1234) + (1 \leftrightarrow 2) =d^{1a4}f^{23}\,_a+d^{13a}f^{24}\,_a+d^{1a2}f^{34}\,_a +(1 \leftrightarrow 2) =0\,.
\eqae
Since the symmetry property of each term is distinct, the constraint is satisfied only if each term is individually zero. Thus imposing unitarity and locality, enforces the group theory factor constraint, 
\eq
d^{abc}f^{de}\,_a=0\, ,
\label{4DConstraint}
\eqe
which is nothing but the anomaly cancellation condition for the non-abelian box anomaly! 

Let us pause and see what we have achieved. Starting with tree-amplitudes, generalized unitarity constructs putative loop-amplitudes that are manifestly unitary. However, they fail to be local. Imposing locality forces inclusion of rational terms to cancel the spurious singularities within these putative loop-amplitudes. Here the chiral theory differs from the non-chiral theories: the requisite rational terms introduce ``new" factorization poles with non-trivial residues. In non-chiral theories, such factorization channels are simply absent. Because there are no suitable tree-amplitudes which it can factorize into, the presence of these residues are simply inconsistent, and therefore must vanish. The only way a chiral theory can achieve this vanishing is with the aid of the group theory factors. This leads precisely to the non-abelian anomaly cancellation condition. Nowhere in the discussion did we invoke or mention of gauge symmetry. Furthermore, as $A_{1/2,\,Cut}^{1-{\rm L, odd}}$ is both IR and UV-finite, it does not have any regulator dependence. Thus the amplitude presents a \textit{manifestly} regulator independent representation of the inconsistencies of chiral theories.

We now move on to higher dimensional chiral gauge and gravity theories, where the story becomes much more interesting. 

\subsection{D=6 QED\label{sec:6DQED}}
We now move on to six-dimensions, and consider the simplest chirally coupled system, chiral QED. Using the six-dimensional spinor helicity formalism introduced by Cheung and O'Connel~\cite{6DSpinor}, the  four-point tree-amplitude of two photons and two chiral-fermions is,
\eq
A_{4}(\gamma_{a_1\dot{a}_1}\gamma_{a_2\dot{a}_2}e_{a_3}e_{a_4})=\frac{\langle 1_{a_1}2_{a_2}3_{a_3}4_{a_4}\rangle[1_{\dot{a}_1}|3|2_{\dot{a}_2}]}{tu}\,,
\eqe
where $(a_i,\dot{a}_i)$ are the SU(2)$\times$ SU(2) little group indices for each external leg. To extract the parity-odd piece contribution from the chiral fermions to the loop, we again take the difference between the chiral and anti-chiral fermion contributions to the loop. For this purpose we will need the amplitude for chiral fermions as well, which is given by:
\eq
A_{4}(\gamma_{a_1\dot{a}_1}\gamma_{a_2\dot{a}_2}e_{\dot{a}_3}e_{\dot{a}_4})=\frac{[1_{\dot{a}_1}2_{\dot{a}_2}3_{\dot{a}_3}4_{\dot{a}_4}]\langle1_{a_1}|3|2_{a_2}\rangle}{tu}\,.
\eqe
We will follow~\cite{YT6D} (see also \cite{Scott6D}) to apply generalized unitarity method in six-dimensions. A brief review of how generalized unitarity method is applied in our context in appendix~\ref{UnitarityAppendix}. The chiral integral coefficients, defined as the difference between chiral and anti-chiral fermion loop, are:
\beqa\label{6DQEDC}
\hspace{-0.2cm}\begin{array}{ccc}
 \displaystyle C_4(1,2,3,4)=\frac{(s-t)}{3u^2}F^{(4)}\,, &\displaystyle C_4(1,3,4,2)=\frac{(u-s)}{3t^2}F^{(4)}, &\displaystyle C_4(1,4,2,3)=\frac{(t-u)}{3s^2}F^{(4)}\,,\\
&&\\
\displaystyle C_{3s}=\frac{4}{3}\left[\frac{1}{u^2}-\frac{1}{t^2}\right]F^{(4)}\,, &\displaystyle C_{3t}=\frac{4}{3}\left[\frac{1}{s^2}-\frac{1}{u^2}\right]F^{(4)}\,,
 &\displaystyle C_{3u}=\frac{4}{3}\left[\frac{1}{t^2}-\frac{1}{s^2}\right]F^{(4)}\,,
\end{array}
\eeqa
where $C_4(1,2,3,4)$ represent the coefficient for the box integral with ordered legs 1,2,3,4, and $C_{3s} (C_{3t})$ represent one-mass $s$-channel ($t$-channel) triangle coefficients. The function $F^{(4)}$ is,
\eqa\label{F4Def}
 F^{(4)}\equiv\frac{1}{2}\big(\langle 4_{d}|p_2p_3|4_{\dot{d}}]F_1\wedge F_2\wedge F_3+ (\sigma_i){\rm cyclic}\big)\,,
\eqae
where $\sigma_i=-1$ for odd cyclic permutations and $+1$ for even. The on-shell form of the wedge product of three field strengths is simply\footnote{The sign between the two terms can be determined from the fact that the parity-odd configuration is symmetric under the exchange of any two labels, while the parity-even configuration is anti-symmetric.}
\eqa
F_1\wedge F_2\wedge F_3 = \left( \langle1_a|2_{\dot{b}}]\langle2_b|3_{\dot{c}}]\langle3_c|1_{\dot{a}}]+\langle2_b|1_{\dot{a}}]\langle1 _a|3_{\dot{c}}]\langle3_c|2_{\dot{b}}] \right)\,. 
\eqae
Thus the cut-constructed part of the amplitude is given by
\eqa\label{6DQEDI}
 A_{1/2,\,Cut}^{1-{\rm L, odd}}&=&\sum_{i,j,k,l\in S_4}C_4(i,j,k,l)I_4(i,j,k,l)+C_{3s}I_{3s}+C_{3t}I_{3t}+C_{3u}I_{3u}\,.
\eqae
Explicitly, the massless-box, one-mass triangle, and (massive) bubble scalar integrals, in $(6-2\epsilon)$-dimnesions, evaluate to:
\eqa \label{Integrals}
\nonumber I_3[K^2]&=&\frac{1}{2 \e} + \frac{1}{2} \left(3 - \gamma_E - \log[K^2]\right)\,,\; \nonumber\\
\nonumber I_2[K^2]&=&-\frac{K^2}{6 \e} + \frac{K^2}{18} \left(-8 +3\gamma_E +3\log[K^2]\right)\,,\\ 
I_4(1,2,3,4)&=&-\frac{\log^2 x+ \pi^2}{2u}\;. \label{Integrals}
\eqae
Combining eq.(\ref{6DQEDC}), eq.(\ref{6DQEDI}), and eq.(\ref{Integrals}), it is easy to see that again there are no UV-divergences in $A_{1/2,\,Cut}^{1-{\rm L, odd}}$, as expected. Note in chiral QED, there is no color-ordering and all propagator poles are physical. But to ensure locality, again the residue of any factorization poles must correspond to the product of lower-point amplitudes. In the vicinity of the $t\rightarrow 0$ pole, the cut-constructed terms approach
\eq
\label{QEDPole}
A_{\frac{1}{2},\,Cut}^{1-{\rm L, odd}}|_{t\rightarrow 0}=\frac{F^{(4)}}{tu}+\mathcal{O}(t^0)\,.
\eqe
We see the presence of a factorization pole with mass-dimension 4 residue. 

We now analyze all possible three-point amplitudes involving at least two vectors with mass-dimension 1,2 and 3.\footnote{We don't consider mass-dimension 0 or 4, since it is easy to see a vector cannot couple with a mass-dimension 0 coupling. The absence of mass-dimension 0 three-point amplitude makes any possible mass-dimension 4 amplitude irrelevant for the purpose of constructing an acceptable residue.} As usual, three-point kinematics are degenerate, and one needs to take special care in defining the on-shell elements. In~\cite{6DSpinor}, special unconstrained variables were introduced to parameterize the kinematics, which carry redundant degrees of freedom. The ``gauge invariance" requirement is then the manifestation of Lorentz invariance for the three-point on-shell amplitudes. A systematic study of all such possible three-point amplitudes involving massless fields were presented in~\cite{YtBartek}, facilitates the following analysis:
\begin{itemize}
  \item (3,1): As with four-dimensions, mass-dimension one three-point amplitudes involving two vectors necessarily imply the third particle is also a vector, and the three-vector amplitude is unique, coming from $F^2$. This then restricts the mass-dimension three-amplitude to be of three vectors as well, which is also unique, coming form an $F_{\mu}\,^{\nu}F_{\nu}\,^{\rho}F_{\rho}\,^{\mu}$ operator. Since the two operators are parity even, they cannot contribute to a parity odd-amplitude.
  \item (2,2): There are three possible mass-dimension two three-point amplitudes involving two vectors: coupling to a scalar, a graviton and a two-form. The former two correspond to the operator $\phi F^2$ and $\sqrt{g} F^2$, which also does not introduce any parity odd contribution. The third possibility is a two-form $B_{\mu\nu}$, which has two possible couplings, $H^2$ and $B\wedge F\wedge F$, where $H_{\mu\nu\rho}\equiv \partial_{[\mu}B_{\nu\rho]}+A_{[\mu}\partial_{\nu}A_{\rho]}$ is the three-form field strength. Note that the $BF^2$ operator indeed introduces a parity-odd contribution. 
 \end{itemize}
Thus we see that the only way for the factorization pole in eq.(\ref{QEDPole}) to have a reasonable residue is to understand it as the propagation of a two-form. In other words, by forcing the unitarity result to also satisfy locality, the scattering amplitude ``responds" by telling us there is a new particle in the spectrum, the two-form of the Green-Schwarz mechanism~\cite{GS}! Note that we again see $A_{\frac{1}{2},\,Cut}^{1-{\rm L, odd}}$ is completely finite, and thus this discussion is manifestly independent of any regulator. 

Later, we will study the residue more closely to see that it is indeed given by the gluing of the $H^2$ and $B\wedge F\wedge F$ interactions. But first we consider chiral-fermions coupled to Yang-Mills, i.e. $6D$ chiral QCD.

\subsection{D=6 QCD\label{sec:6DQCD}}
We now consider a chiral fermion coupled to Yang-Mills theory. The relevant four-point amplitudes for two gluon and two chiral (anti-chiral) fermions are:
\eqa
{\rm Chiral}:\quad A_{4}(g_{a_1\dot{a}_1}g_{a_2\dot{a}_2}q_{a_3}q_{a_4})&=&\frac{\langle 1_{a_1}2_{a_2}3_{a_3}4_{a_4}\rangle[1_{\dot{a}_1}|3|2_{\dot{a}_2}]}{ts}\,,\\
{\rm Anti-chiral}:\quad A_{4}(g_{a_1\dot{a}_1}g_{a_2\dot{a}_2}q_{\dot{a}_3}q_{\dot{a}_4})&=&\frac{[1_{\dot{a}_1}2_{\dot{a}_2}3_{\dot{a}_3}4_{\dot{a}_4}]\langle1_{a_1}|3|2_{a_2}\rangle}{ts}\,.
\eqae 
The integral coefficients now become
\eqa
\nonumber C_4&=&\frac{(s-t)}{6u^2}F^{(4)}\,,\;\; C_{3s}=-\frac{(s-t)}{6tu^2}F^{(4)}\,,\\
 C_{3t}&=&-\frac{(s-t)}{6su^2}F^{(4)}\,,\;\; C_{2s}=\frac{F^{(4)}}{stu}\,,\;\;C_{2t}=-\frac{F^{(4)}}{stu} \,,
\eqae
where $F^{(4)}$ is defined as before. Note that the bubble and triangle coefficients are such that there is no overall UV-divergence.  Indeed one funds that the integrated result for the cut-constructible piece is given by:
\eq
\label{QCDPole}
A_{\frac{1}{2},\,Cut}^{1-{\rm L, odd}}=F^{(4)}\left[\frac{(t-s)(\pi^2+\log^2[x])}{12u^3}+\frac{\log[x]}{3u^2}+\frac{s-t}{18stu}\right]\,.
\eqe
Again the ubiquitous $u$-poles in the planar amplitude requires us to inquire if it has a rational residue. One finds:
\eq
\label{QCDPole}
A_{\frac{1}{2},\,Cut}^{1-{\rm L, odd}}\Big|_{u\rightarrow 0}=-\frac{F^{(4)}}{18tu}+\mathcal{O}(u^0)\,.
\eqe
Note that, un-like in four-dimensions where we can immediately conclude that one needs to include rational term, here we need to consider the fact that the same pole will appear in other color-orderings. In other words the presence of this pole may cancel between different color-orderings, just like the physical poles introduced by the rational term in 4D chiral QCD can cancel among different color-orderings, as discussed in sec.~\ref{sec:4DChiralQCD}.  Indeed the same pole also appears in other color-ordering. Analyzing the behavior of the cut constructed color-dressed amplitude, one finds the following collinear behavior in the $s$-,$t$- and $u$-channels:
\beqa\label{QCDFac}
\hspace{-0.3cm}&&\mathcal{A}_{\frac{1}{2},\,Cut}^{1-{\rm L, odd}}~\underrightarrow{\;\;u=0\;\;}~-\frac{F^{(4)}}{18ut}str(T^4)+\mathcal{O}(u^0)\,,~~  \mathcal{A}_{\frac{1}{2},\,Cut}^{1-{\rm L, odd}}~\underrightarrow{\;\;s=0\;\;}~-\frac{F^{(4)}}{18su}str(T^4)+\mathcal{O}(s^0)\,,\nonumber\\
&&\mathcal{A}_{\frac{1}{2},\,Cut}^{1-{\rm L, odd}}~\underrightarrow{\;\;t=0\;\;}~-\frac{F^{(4)}}{18ts}str(T^4)+\mathcal{O}(t^0)\,.
\eeqa  
Here $str(T^4)$ represents the symmetric trace of four generators. Thus we see that the collinear factorization pole will not contribute, if the group theory structure satisfies: 
\eq\label{6DanomCond}
str(T^4)=0\,.
\eqe
This is nothing but the anomaly free condition in six-dimensions. Note that the reason why spurious poles of the cut-constructed answer can cancel between different ordering can attributed to the fact that in six-dimensions, the cut-constructed answer actually already contains a rational term, whereas such terms are not present in four-dimensions. The distinction is that in six-dimensions there are two scalar integrals in eq.\eqref{Integrals} that contain UV-divergences. The relative coefficients between the $1/\epsilon$-divergent pieces and finite pieces are different between the two integrals. Thus if the final answer is UV-finite, then when combined with cyclic symmetry one can conclude there must be a left-over rational piece already present in the cut-constructible terms. In $D=4$ there is only one UV-divergent scalar integral, the bubble. Thus if the UV-divergence cancels in $D = 4$, so will the rational term in the scalar bubble. 

On the other hand, if $str(T^4)\neq0$, in $D = 6$ chiral QCD, the cut-constructible part has non-trivial rational residues on each factorization channel, and locality then requires each such residue to correspond to the exchange of a particle. From the analysis of QED, we know that this pole can only be interpreted as an exchange of a GS two-form. For this interpretation to be true the group theory factor must factorize, i.e. we must have $str(T^4)\sim tr(t^2)tr(t^2)$ for some representation $t$. This is the well known story that when the generators are rewritten in fundamental representation, for a general gauge group one has:
\eq
str(T^4)=\alpha\,str(t^4)+\beta\,tr(t^2)tr(t^2)\,.
\eqe
For the poles to have a physical interpretation we must have $\alpha=0$. 

However even with $\alpha=0$, and the symmetric trace factorizes, there is still a problem. Eq.(\ref{QCDFac}) becomes,
\eq
\mathcal{A}_{\frac{1}{2},\,Cut}^{1-{\rm L, O}}\;\;\underrightarrow{\;\;u=0\;\;}\;\;-\frac{F^{(4)}}{18ut}\left[tr(t_1t_2)(t_3t_4)+tr(t_1t_3)(t_2t_4)+tr(t_1t_4)(t_3t_2)\right)+\mathcal{O}(u^0)]\, .
\eqe
Clearly, only the group theory factor $tr(t_1t_3)(t_2t_4)$ makes any sense as a factorization channel for the $u$- channel pole. In other words, this $u$-pole reside should not have any term proportional to $tr(t_1t_2)(t_3t_4)$ or $tr(t_1t_4)(t_2t_3)$. Similar conclusions can be reached from the $s$- and $t$-pole analysis. Thus for the poles to have an interpretation as a proper factorization channel, extra rational terms must be added to the cut-constructible terms. From symmetry properties required from the trace structure, we can deduce that the requisite rational term is:
\eq
\mathcal{R}=F^{(4)} \left[tr(t_1t_2)(t_3t_4)\frac{u-t}{18stu}+tr(t_1t_3)(t_2t_4)\frac{t-s}{18stu}+tr(t_1t_4)(t_3t_2)\frac{s-u}{18stu}\right]\,.
\label{FinalR}
\eqe
Indeed we now find that with the above rational term, we now have:
\eq\label{ChiralQCDBehave}
\begin{array}{cc}\; & \underrightarrow{\;\;u=0\;\;}\;\;-\frac{F^{(4)}}{6ut}tr(t_1t_3)(t_2t_4)+\mathcal{O}(u^0) \\ \mathcal{A}_{\frac{1}{2},\,Cut}^{1-{\rm L, O}}+\mathcal{R} & \underrightarrow{\;\;s=0\;\;}\;\;-\frac{F^{(4)}}{6su}tr(t_1t_2)(t_3t_4)+\mathcal{O}(s^0) \\ \; & \underrightarrow{\;\;t=0\;\;}\;\;-\frac{F^{(4)}}{6ts}tr(t_1t_4)(t_3t_2)+\mathcal{O}(t^0)\end{array}\,.
\eqe

In summary, we have found that for chiral fermions coupled to Yang-Mills in six-dimensions, the answer obtained from unitarity methods can be local only if $str(T^4)=0$ or $str(T^4)=\beta\,tr(t^2)tr(t^2)$, for some representation $t$. Furthermore, for the case that $str(T^4)\neq 0$, locality requires one to add a rational term $\mathcal{R}$ given in eq.(\ref{FinalR}). With this additional term one obtains a unitary and local answer. One might wonder what is this magical rational term $\mathcal{R}$? Since it is constructed from on-shell methods, it is gauge invariant. In sec~\ref{sec:Feynman}, we will show that, in terms of Feynman diagrams, this is nothing but the gauge invariant combination of the anomalous rational term plus the tree-diagrams in the GS mechanism! But, before moving on to Feynman diagrams, we analyze whether or not the residue is indeed given by the gluing of two 3-point amplitudes involving the exchange of a two-form.

\section{Factorization properties of the rational term}
In this section we study the residue of the single pole that appeared both in the chiral QED, in~eq.(\ref{QEDPole}), and chiral QCD, in~eq.(\ref{ChiralQCDBehave}). They are all proportional to $F^{(4)}$ defined in~\ref{F4Def}, divided by the Mandlestam variable of the non-factorizing channel. Dimensional and little group analysis tell us that the residues on these poles can only correspond to propagation of a two-form. However we have not explicitly provided the three-point amplitudes whose gluing should in principle give the this residue. Three-point elements are highly constrained, and not surprisingly, the three-point amplitude for two vector fields and a two-form is unique~\cite{YtBartek}. In this section, we will show that the parity odd part of the two-form  exchange indeed gives the residue in eq.(\ref{QEDPole}) and eq.(\ref{ChiralQCDBehave}).
.

\subsection{Review on six-dimensional On-shell variables for three-point kinematics\label{sec:3ptKinematics}}
The possible three-point amplitude in six-dimensions have been studied extensively and is severely constrained using the following SU(2) spinors~\cite{6DSpinor}:
\eqa
\nonumber&&\langle 1_{a}|2_{\dot{a}}]\equiv u_{1a}\tilde{u}_{2\dot{a}}\,,\quad\langle 2_{a}|1_{\dot{a}}]\equiv -u_{2a}\tilde{u}_{1\dot{a}}\,,\\
\nonumber&&\langle 2_{a}|3_{\dot{a}}]\equiv u_{2a}\tilde{u}_{3\dot{a}}\,,\quad\langle 3_{a}|2_{\dot{a}}]\equiv -u_{3a}\tilde{u}_{2\dot{a}}\,,\\
&&\langle 3_{a}|1_{\dot{a}}]\equiv u_{3a}\tilde{u}_{1\dot{a}},\quad\langle 1_{a}|3_{\dot{a}}]\equiv -u_{1a}\tilde{u}_{3\dot{a}}\,,
\eqae
as well as their pseudo inverse:
\eq
u_{ia}w_{ib}-u_{ib}w_{ia}=\e_{ab},\quad\tilde{u}_{i\dot{a}}\tilde{w}_{i\dot{b}}-\tilde{u}_{i\dot{b}}\tilde{w}_{i\dot{a}}=\e_{\dot{a}\dot{b}}\,.
\eqe
Note that the above definitions are invariant under the following rescaling and shift symmetries:
\eqa\label{uwsymmetries}
(u_{i} ,w_{i})\rightarrow (\alpha u_{i} ,\alpha^{-1}w_{i}),&&\quad(\tilde{u}_{i} ,\tilde{w}_{i})\rightarrow (\alpha^{-1} \tilde{u}_{i} ,\alpha\tilde{w}_{i}),\\
w_{i}\rightarrow w_{i}+b_{i}u_{i}, &&\quad \tilde{w}_{i}\rightarrow \tilde{w}_{i}+\tilde{b}_{i}\tilde{u}_{i} \, .
\eqae 
Repeated indices are not summed. Since these symmetries arise from their definition with respect to Lorentz invariants, preservation of these symmetries is equivalent to Lorentz invariance. Thus all possible Lorentz invariant three-point amplitudes must be a polynomial of these objects satisfying the rescaling and shift symmetry. Momentum conservation further implies 
\eq
u_{ia}|i^a\rangle=u_{ja}|j^a\rangle,\quad \tilde{u}_{i\dot{a}}|i^{\dot{a}}]=\tilde{u}_{j\dot{a}}|j^{\dot{a}}],\quad\sum_{i}w_{ia}|i^a\rangle= \sum_{j}\tilde{w}_{j\dot{a}}|j^{\dot{a}}]=0\,.
\eqe
Again repeated indices are not summed, and when combined with eq.(\ref{uwsymmetries}) implies that $b_1+b_2+b_3=0$, $\tilde{b}_1+\tilde{b}_2+\tilde{b}_3=0$.

The three-point amplitudes involving one (anti-) chiral two-form and two vector fields are uniquely fixed by Lorentz invariance and little group constraint~\cite{YtBartek}:
\eqa
\nonumber&&\;A_{(ad),b\dot{b},c\dot{c}}^{B}=\bigg(\Delta_{(a|bc}u_{1|d)}\bigg)\tilde{u}_{2\dot{b}}\tilde{u}_{3\dot{c}}\,,\;\;A_{(\dot{a}\dot{d}),b\dot{b},c\dot{c}}^{\bar{B}}=u_{2b}u_{3c}\bigg(\tilde{\Delta}_{(\dot{a}|\dot{b}\dot{c}}\tilde{u}_{1|\dot{d})}\bigg)\,,
\label{3pt}
\eqae
where
\eqa
\nonumber\Delta_{abc}&=&(u_{1a}u_{2b}w_{3c}+u_{2b}u_{3c}w_{1a}+u_{3c}u_{1a}w_{2b})\,,\\
\tilde\Delta_{\dot{a}\dot{b}\dot{c}}&=&(\tilde{u}_{1\dot{a}}\tilde{u}_{2\dot{b}}\tilde{w}_{3\dot{c}}+\tilde{u}_{2\dot{b}}\tilde{u}_{3\dot{c}}\tilde{w}_{1\dot{a}}+\tilde{u}_{3\dot{c}}\tilde{u}_{1\dot{a}}\tilde{w}_{2\dot{b}})\,.
\eqae
The little group indices tells us that the two vector fields sit on legs 2 and 3, whilst the field on leg $1$ transform as a (3,0) and (0,3) in $A^{B}$ and $A^{\bar{B}}$ respectively. Because the six on-shell components of a two-form can be split into self-dual and anti-self dual sectors, $A^{B}$ and $A^{\bar{B}}$ represents the couplings of self-dual anti-self-dual two-forms, respectively. 

\subsection{Gluing $(A^{B},A^{B})$ and $(A^{\bar{B}},A^{\bar{B}})$}
We now consider the gluings of two $A^{B}$s, as well as of $A^{\bar{B}}$s. Taking the difference of these two products yields the parity-odd contribution from the two-form exchange.\footnote{Recall that the coupling of two-forms to gauge field is done via the three-form field strength $H^2$ as well as $B\wedge F\wedge F$. Denoting the resulting three-point coupling as $V_1$ and $V_2$ respectively, we are interested in the parity-odd contribution from the gluing of $V_1$ and $V_2$. We will discuss this in more detail in subsection~\ref{sec:TwoFormTree} } Let us consider the product of three-point amplitudes in a $t$-channel exchange. The kinematic labels are as follows:
$$\includegraphics[scale=0.65]{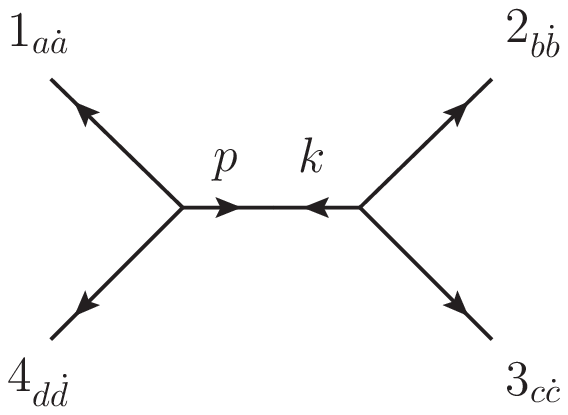}~~~~~~~.$$
The sewing of two self-dual tensors and anti-self-dual tensors are given by:\footnote{For the variables related to legs $k$ and $p$, one can show that $(u_p\cdot u_k)(\tilde{u}_{k}\cdot \tilde{u}_{p})=-s$. Using the $b_i$ shift invariance, one can "gauge fix" $w_p$ and $w_k$ such that:
$$(w_p\cdot u_p)=(w_k\cdot u_k)=0$$
and similar result for $(\tilde{w}_p\cdot \tilde{u}_p)$ and $(\tilde{w}_k\cdot \tilde{u}_k)$. Furthermore, we also have $(w_p\cdot w_k)=1/(u_p\cdot u_k)$. The leftover rescaling symmetries $\alpha_L$ can $\alpha_R$ can be further fixed by requiring $(u_k\cdot u_p)=(\tilde{u}_{k}\cdot \tilde{u}_{p})=\sqrt{-s}$.}
\eqa
\nonumber A_{3L}^{B}A_{3R}^{B}&=&\frac{1}{2}\left(\Delta_{(e|da}u_{p|f)}\Delta^{e}\,_{bc}u_{k}^f\right)\left(\tilde{u}_{1\dot{a}}\tilde{u}_{4\dot{d}}\tilde{u}_{2\dot{b}}\tilde{u}_{3\dot{c}}\right)\\
\nonumber&=&\frac{1}{2}\Big[u_{1a}u_{4d}u_{2b}u_{3c}-2s(u_{1a}w_{4d}+u_{4d}w_{1a})(u_{2b}w_{3c}+u_{3c}w_{2b})\Big]\tilde{u}_{1\dot{a}}\tilde{u}_{4\dot{d}}\tilde{u}_{2\dot{b}}\tilde{u}_{3\dot{c}}\\
\nonumber A_{3L}^{\bar{B}}A_{3R}^{\bar{B}}&=&\frac{1}{2}u_{1a}u_{4d}u_{2b}u_{3c}\Big[\tilde{u}_{1\dot{a}}\tilde{u}_{4\dot{d}}\tilde{u}_{2\dot{b}}\tilde{u}_{3\dot{c}}-2s(\tilde{u}_{1\dot{a}}\tilde{w}_{4\dot{d}}+\tilde{u}_{4\dot{d}}\tilde{w}_{1\dot{a}})(\tilde{u}_{2\dot{b}}\tilde{w}_{3\dot{c}}+\tilde{u}_{3\dot{c}}\tilde{w}_{2\dot{b}})\Big]\,,
\label{3times3}
\eqae
where $k$ and $p$ refers to the momentum in the internal leg with $k=-p$. Thus the parity-odd contribution to the rational term coming from a two-form exchange is given by:
\eqa
\nonumber Res_{t}&=& A_{3L}^{B}A_{3R}^{B}-A_{3L}^{\bar B}A_{3R}^{\bar B}\\
\nonumber &=&s(u_{1a}w_{4d}+u_{4d}w_{1a})(u_{2b}w_{3c}+u_{3c}w_{2b})(\tilde{u}_{1\dot{a}}\tilde{u}_{4\dot{d}}\tilde{u}_{2\dot{b}}\tilde{u}_{3\dot{c}})\\
&&-s(\tilde{u}_{1\dot{a}}\tilde{w}_{4\dot{d}}+\tilde{u}_{4\dot{d}}\tilde{w}_{1\dot{a}})(\tilde{u}_{2\dot{b}}\tilde{w}_{3\dot{c}}+\tilde{u}_{3\dot{c}}\tilde{w}_{2\dot{b}})(u_{1a}u_{4d}u_{2b}u_{3c})\,.
\label{t-residue}
\eqae
Following~\cite{6DSpinor} one can straightforwardly rewrite this in terms of the full six-dimensional spinors as:
\eqa
Res_{t}&=&-\frac{\langle1_a2_b3_c4_d\rangle[1_{\dot{a}}|2|3_{\dot{c}}][4_{\dot{d}}|1|2_{\dot{b}}]-\langle 1_a|2|3_c\rangle\langle4_d|1|2_b\rangle[1_{\dot{a}}2_{\dot{b}}3_{\dot{c}}4_{\dot{d}}]}{s}\,.
\label{t-Ansatz}
\eqae

We will now show that eq.(\ref{t-Ansatz}) is equivalent to $F^{(4)}/s$, the residue in~eq.(\ref{QEDPole}) and~eq.(\ref{ChiralQCDBehave}). First we list the following useful identities:
\beqa
\label{MysteryId}
\langle ijml\rangle[i|j|n]
\notag&=&\langle i|j|m\rangle\big\{\langle j|n]\langle l|i]-\langle j|i]\langle l|n]\big\}\\
\notag &&+\langle i|j|l\rangle\big\{\langle j|i]\langle m|n]-\langle j|n]\langle m|i]\big\}+\langle l|j|m\rangle \langle j|i]\langle i|n]\,,
\\
\notag \left[ijml\right] \langle i|j|n\rangle&=&[i|j|m]\big\{[\langle n| j] \langle i|l]-\langle i|j] \langle n|l]\big\}\\
&&+[ i|j|l]\big\{[ \langle i|j]\langle n|m]-\langle n|j]\langle i|m]\big\}+[ l|j|m] \langle i|j]\langle n|i]\,.
\eeqa
We have suppressed the little group labels. These identities can be derived by noting that each product on the LHS involves the product of two SU(4) Levi-Cevitas, the latter of which can be rewritten as a product of Kronecker deltas. Defining 
\eq
\label{fM}f_{ijml}\equiv \langle ijml\rangle[i|j|m][l|i|j]-[ijml]\langle i|j|m\rangle\langle l|i|j\rangle\,,
\eqe 
and using momentum conservation, $k_i+k_j+k_m+k_l=0$, it is trivial to show that $f_{ijml}=f_{jilm}=f_{mlij}=f_{lmji}$.
Substituting eq.(\ref{MysteryId}) into eq.\eqref{fM}, one finds
\beqa
f_{ijml}&=&s_{ij}\big\{\langle i|m] \langle m|j]\langle j|l]\langle l|i]+\langle i|j]\langle j|l]\langle l|m]\langle m|i]
\notag\\
\notag &&~~~~~-\langle i|l]\langle l|j]\langle j|m]\langle m|i] -\langle i|m]\langle m|l]\langle l|j]\langle j|i] \big\}\\
\notag &&+\big\{\langle i |j]\langle j|i]\langle l|m]-\langle l|i]\langle i |j]\langle j|m]-\langle l |j]\langle j|i]\langle i|m]\big\}\langle m|p_ip_j|l]\\
\notag &&-\big\{\langle i|j]\langle j |i]\langle m|l]-\langle m|j]\langle j |i]\langle i|l]-\langle m |i]\langle i|j]\langle j|l]\big\}\langle l|p_jp_i|m]\\
 &&+2\big\{\langle i |j]\langle j|m]\langle m|i]  +\langle i|m]\langle m |j]\langle j|i] \big\}\langle l|p_ip_j|l] 
\eeqa
 so that one can show that
\eqa
&&8F^{(4)}=f_{1234}+f_{2143}+f_{3412}+f_{4321}-f_{2134}-f_{1243}-f_{3421}-f_{4312}\,.
\eqae
Thus we conclude that 
\eq
2F^{(4)}=f_{1234}-f_{2134}\,.
\eqe
Finally, one can show that $f_{ijml}=-f_{jiml}$ by using Schouten identity, 
\eqa
\nonumber [i|j|m][l|i|j]&=&2s_{ij}[mlij]-[l|j|i][m|i|j]\,,\\
\langle i|j|m\rangle\langle l|i|j\rangle& =&2s_{ij}\langle mlij\rangle-\langle l|j|i\rangle\langle m|i|j\rangle\,.
\label{S1S22}
\eqae
Thus in conclusion, we indeed find that 
\eq
Res_{t}=-\frac{f_{1234}}{s}= -\frac{F^{(4)}}{s}\,,
\label{t-Ansatz2}
\eqe
which is precisely the residue observed in the $t$-channel factorization limit in chiral QED and QCD, i.e. eq.(\ref{QEDPole}) and eq.(\ref{ChiralQCDBehave}). This confirms that the final parity-odd amplitude that satisfies unitarity and locality has a factorization channel which is consistent with the tree-level exchange of a two-form. 
\subsection{Comments on tree-level gauge invariance\label{sec:TwoFormTree} }
Let us consider the following question: is it possible to construct a parity-odd four-point gluon amplitude that is purely rational and whose factorization channels include the exchange of a self-dual two-form? From eq.(\ref{t-Ansatz2}), we see that the $t$-channel exchange of a two-form has a residue that also involves an $s$ ($u$) channel pole. This implies that any amplitude that has a $t$-channel $B$-field exchange must also include an $s$- or an $u$-channel exchange.\footnote{For $t=0$, $s=-u$. So on the $t$-channel pole one cannot distinguish an $s$-channel from an $u$-channel pole.} Without loss of generality, let us assume that the other channel is an $s$-channel pole. Since the residue on the $s$-channel pole is simply a cyclic rotation of eq.(\ref{t-Ansatz2}), it is easy to see that the only solution is given by:\footnote{It is impossible to have terms that do not have poles. This is simplest to see using polarization vectors. The amplitude has mass-dimensions 2 and involves a Levi-Cevita. The only pole-free structure that satisfies this criteria is that of the form $\epsilon(F_1F_2e_3e_4)$, which cannot be made gauge invariant.} 
\eq
\label{GSTreeBS}
\frac{F^{(4)}(s-t)}{stu}\,.
\eqe 
Thus, naively, one would conclude that the answer to the question at the beginning of this section is yes. Note however, that the result in eq.(\ref{GSTreeBS}) has a non-vanishing $u$-pole, whose residue is \textit{twice} that of the would-be two-form exchange residues in the $s$- or $t$-channels:
\eq
Res_u\left[\frac{F^{(4)}(s-t)}{stu}\right]=\frac{-2F^{(4)}}{s},\; Res_s\left[\frac{F^{(4)}(s-t)}{stu}\right]=\frac{-F^{(4)}}{u},\; Res_t\left[\frac{F^{(4)}(s-t)}{stu}\right]=\frac{F^{(4)}}{u}\,.
\eqe
This factor of $2$ makes it impossible for the collinear limit of all channels in eq.(\ref{GSTreeBS}) to consistently factorize into the same three-point amplitudes. In other words, one cannot obtain a purely rational four-point parity-odd amplitude whose poles can be consistently interpreted as the exchange of a two-form. 

The difficulty above is precisely an on-shell way of saying that one cannot have consistent tree-level coupling of a two-form with vector fields, if the coupling allows for parity-odd amplitudes. This is a reflection of the fact that one cannot maintain gauge invariance if the vectors couple to the  two-form via both the three-form field strength, and a parity-odd coupling. Recall that in the standard Green-Schwarz mechanism, one introduces an irrelevant operator $B\wedge F\wedge F$, which when combined with the cross terms in $H^2\equiv (dB+\omega_3)^2$ with $\omega_3=tr(FA-A^3/3)$, generates the requisite gauge anomaly. The anomaly arises from the fact that one cannot define the gauge variation of the two-form $B_{\mu\nu}$ such that \textit{both} $B\wedge F\wedge F$ and $H^2$ are simultaneously gauge invariant. As a consequence, the tree-diagram that involves a vertex from $B\wedge F\wedge F$ and another from $H^2$ is always anomalous under the gauge variation:
\eq
\left.\;\left(\vcenter{\hbox{  \includegraphics[scale=0.6]{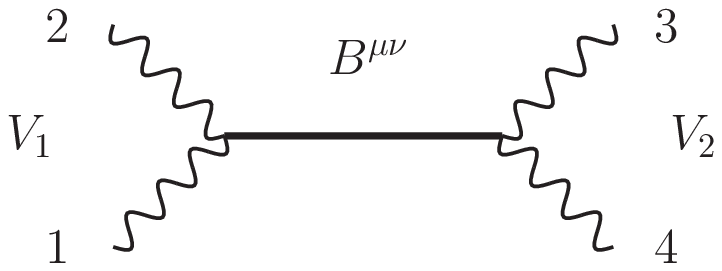} }}\right)\right|_{\epsilon_1\rightarrow k_1}\, =2 F_2\wedge F_3\wedge F_4
\eqe
where $V_1=\partial_{[\mu}B_{\nu\rho]}tr(A^\mu \partial ^\nu A^\rho)$ and $V_2=B\wedge tr(F\wedge F)$. Note that the combination of $V_1$ and $V_2$ is precisely what produces a parity odd-amplitude. 

The only way out of this predicament, from an on-shell view point, is if something else absorbed the excess residues. This is precisely what the cut-constructible part of the one-loop chiral fermion accomplished in sec.~\ref{sec:6DQCD}! Thus we have arrived at the following conclusion: A tree-level exchange of two-forms coupled to external gauge fields is consistent with locality and unitarity \textit{only if} it is accompanied by the cut-constructible part of a chiral fermion loop. Thus instead of gauge anomaly cancellation, from the on-shell view point it is the cancellation of excess residues on factorization poles. Not surprisingly, eq.(\ref{GSTreeBS}) appears as the rational term required by locality in eq.(\ref{FinalR}). Note that for chiral chiral QED, the same analysis applies, and the only difference is that the rational term cancels in the end, as can be seen from eq.(\ref{FinalR}) by identifying all trace factors.

\section{Gauge invariant rational terms from Feynman rules\label{sec:Feynman}}
In this section, we make contact between the traditional discussion of perturbative anomalies, from analysis of Feynman diagrams, and the previous on-shell presentation. Straightforward analysis of one-loop Feynman diagrams shows that, in $D = 2(n -1)$-dimensions, the first non-zero parity-odd terms appear at $n$-points. As such, only the parity-odd terms in the $n$-gon Feynman diagram are relevant to our discussion. 

After integral reduction, the scalar integral coefficients exactly match those obtained from generalized unitarity. However, rational terms obtained from this integral reduction are not gauge-invariant. This is the source of the anomaly. Crucially, the Green-Schwarz mechanism introduces new gauge-\emph{variant} interactions directly into the action which cancel the gauge-variation of the rational term from the chiral fermion loop. Combining the Green-Schwarz trees with the anomalous rational term remarkably, though unsurprisingly, produces precisely the gauge-invariant rational term, in section~\ref{sec:6DQCD}, required by locality from generalized unitarity.

The calculation is organized as follows. First, we extract the gauge-variant rational terms from the $n = D/2 +1$-point one-loop Feynman diagram with internal chiral fermions and external gluons. The rational term, for a given cyclic ordering of gluons, is given by the product of a permutation-invariant function of the external data multiplied by a color-trace. Second, guided by the color-structure of the full one-loop amplitude, we extract the unique, gauge-invariant, combination of one-loop chiral QCD amplitudes and tree-level, parity-violating, gravitational amplitudes. Finally, we present explicit forms for the $n$-gon integral coefficients, in both six- and eight-dimensions. The structure of these integral coefficients makes it clear that the structure of the highest-order integral coefficient in $D$-dimensions contains all the information concerning the Green-Schwarz mechanism that is conventionally accessed only through an action point of view.

\subsection{The anomalous rational term}
To cleanly extract the parity-odd rational terms in $D=2(n-1)$-dimensions, it is convenient to separate the loop-momenta, $\ell^\mu$, into $D$-dimensional and $-2\epsilon$ dimensional pieces, respectively $\bar{\ell}^\mu$ and $\mu^{\hat{\mu}}$:
\eq
\ell^\mu\rightarrow \bar{\ell}^\mu+\mu^{\hat{\mu}} \, .
\eqe
In this notation, the external momenta is purely $D$-dimensional, and one has, 
\eq
\frac{1}{(\ell+k)^2}=\frac{1}{(\bar\ell+k_1)^2+\mu^2}\,.
\eqe 
Similar separation is required for the definition of the corresponding Gamma matrices. We follow the conventions of 't Hooft and Veltman:
\eq
\gamma^\mu\rightarrow\bar{\gamma}^\mu+\hat{\gamma}^{\hat\mu} \, .
\eqe
The extra $-2\epsilon$-dimensional piece, $\hat{\gamma}^{\hat\mu}$, satisfies the following commutation relations, 
\eq
\{\hat{\gamma}^{\hat\mu},\hat{\gamma}^{\hat\nu}\}=\eta^{\hat\mu\hat\nu},\;\{\hat{\gamma}^{\hat\mu},\bar{\gamma}^{\nu}\}=0,\;\;[\gamma_{-1},\hat{\gamma}^{\hat{\nu}}]=0 \, ,
\eqe
where $\gamma_{-1}$ denotes the $D$-dimensional chiral matrix ($\gamma_{-1}=\gamma_5$ in four-dimensional notation). Note that  $\hat{\gamma}^{\hat\mu}$ may only be contracted with $\mu^{\hat{\mu}}$; all other vectors in the problem are purely $D$-dimensional.

First note that due to the following identity, 
\eq
(1+\gamma_{-1})(\displaystyle{\not}K_{(D)})(1-\gamma_{-1})=2(1+\gamma_{-1})(\displaystyle{\not}K_{(D)}),\;\;(1+\gamma_{-1})(\displaystyle{\not}\mu)(1-\gamma_{-1})=0 \, ,
\eqe
where $K_D$ is some $D$-dimensional vector, it is straightforward to see that the spinor traces of the $n=D/2+1$-point amplitude will project out all $-2\epsilon$-dimensional components of the loop-momenta. For example, in $D=6$ the numerator of the box-diagram $(a)$, derived by stringing together four chiral-fermion-gluon vertex $V_3=\bar{\psi}\displaystyle{\not}A(1+\gamma_{-1})\psi$, is given as:
$$\includegraphics[scale=0.7]{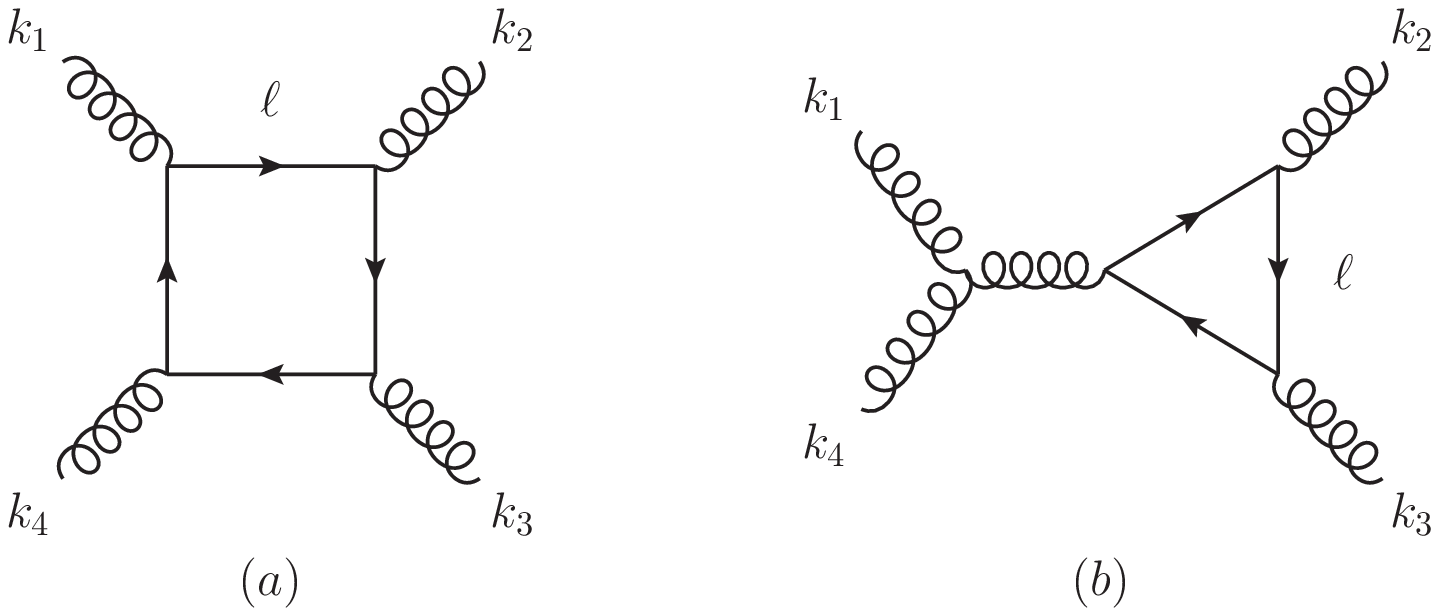}$$
\eqa
\label{na}
\nonumber n_{(a)}&=&tr[\displaystyle{\not}e_1(1+\gamma_{-1})\displaystyle{\not\,}\ell\displaystyle{\not}e_2(1+\gamma_{-1})(\displaystyle{\not\,}\ell-\displaystyle{\not}k_2)\displaystyle{\not}e_3(1+\gamma_{-1})(\displaystyle{\not\,}\ell-\displaystyle{\not}k_2-\displaystyle{\not}k_3)\displaystyle{\not}e_4(1+\gamma_{-1})(\displaystyle{\not\,}\ell+\displaystyle{\not}k_1)]\\ 
&=&8tr[\gamma_{-1}\displaystyle{\not}e_1\displaystyle{\not\,}\bar{\ell}_{(6)}\displaystyle{\not}e_2(1+\gamma_{-1})(\displaystyle{\not\,}\bar{l}-\displaystyle{\not}k_2)\displaystyle{\not}e_3(\displaystyle{\not\,}\bar{\ell}-\displaystyle{\not}k_2-\displaystyle{\not}k_3)\displaystyle{\not}e_4(\displaystyle{\not\,}\bar{\ell}+\displaystyle{\not}k_1)] \, .
\eqae
To reiterate, by virtue of the commutation-relations for $\hat{\gamma}^{\hat{\mu}}$, the loop-momenta in the numerator are purely $D$-dimensional. Similarly, the numerator of the triangle diagram is purely $D$-dimensional:  
\eqa
\nonumber n_{(b)}&=&tr[\gamma_{-1}\displaystyle{\not}J_{1,4}(\displaystyle{\not\,}\bar{\ell}+\displaystyle{\not\,}k_{2})\displaystyle{\not}\e_2\displaystyle{\not\,}\bar{\ell}\displaystyle{\not}\e_3(\displaystyle{\not\,}\bar{\ell}-\displaystyle{\not}k_3)]\\
&=&\epsilon(J_{14}k_{2}e_2\bar{\ell}e_3k_3)\,,
\eqae
where $J_{14}^\mu$ is the current at the three-point gluon vertex. This diagram integrates to zero due to its linear dependence in loop-momentum: through integral reduction, $\bar{\ell}$ is expanded on the vector $k_2$ and $k_3$, and therefore vanishes in the six-dimensional Levi-Cevita tensor. 

Equipped with this specific six-dimensional example, it is straightforward to generalize: the parity-odd amplitudes in $D$-dimensions first appear at $n=D/2+1$-points, and all non-trivial contributions come from the numerator of the $n$-gon chiral fermion loop. Standard gamma-matrix trace identities dictate that the gamma-trace tensor, schematically, reduces in the following way:
\eq
tr[\gamma_{-1}\gamma^{a_1}\cdots \gamma^{a_{2n}}]=\eta^{a_1a_2}\epsilon^{a_3\cdots a_{2n}}+\cdots \, .
\eqe
Again, by the antisymmetry of the $D$-dimensional $\epsilon^{a_1 \cdots a_{D}}$-tensor, no term in the numerator may contain more than three powers of $\bar{\ell}$. Now, integral reduction of the $n$-gon numerator casts the parity-odd amplitude onto a scalar integral basis. With at most three powers of loop-momenta in the numerator, the amplitude is a combination of $n$, $(n-1)$ and $(n-2)$-gon scalar integrals---and the attendant rational terms.\footnote{Note the distinction between $m<n$-gons in the chiral gauge theory and those, obtained by integral reduction \emph{from} the defining $n$-gon, that multiply the \emph{scalar} $m < n$-gon loop-integrals.} Of these three scalar integrals, only the $(n-1)$-gon and the $(n-2)$-gon integrals diverge in the ultraviolet. Their integral coefficients, which are gauge invariant, must ensure cancellation of ultraviolet divergences. 

Since these scalar integral coefficients may be uniquely determined from unitarity cuts, which are simply products of gauge-invariant quantitaties, they must be gauge invariant. However, the same cannot be said of rational terms. Indeed, integral reduction produces rational terms that are not gauge invariant. This is most simply exhibited by terms in the $n$-gon integral with tensor numerators of degree-two in the $D$-dimensional loop-momenta:
\eq
\label{LTest}
\frac{\bar{\ell}^\mu \bar{\ell}^\nu}{\ell^2(\ell-k_1)^2\cdots (\ell+k_n)^2}\,.
\eqe
In integral reduction, one expands 
\eq
\label{Lexp}
\bar{\ell}^\mu \bar{\ell}^\nu= a_0\eta^{\mu\nu} +a_{ij}k_i^\mu k_j^\mu\, \, ,
\eqe
where $i,j$ runs over the independent momenta of the external states. Contraction of both sides of the equation with $\eta^{\mu\nu}$ partially fixes the coefficients $a_0$, and $a_{ij}$. Under this operation, the LHS simply becomes $\bar{\ell}^2=\ell^2-\mu^2$, i.e. an inverse propagator and a $-\mu^2$-term. Insertion of the expansion in eq.(\ref{Lexp}) back into eq.(\ref{LTest}), thus yields a collection of $n$-, $(n-1)$-, and $(n-2)$-gon scalar integrals \textit{and} an $n$-gon scalar integral with a $\mu^2$ numerator. Explicit integration of the latter either vanishes at $\mathcal{O}(\epsilon)$, or gives a rational term. For example, one finds:
\eq
\int \frac{d\ell^{(D)}d\mu^{-2\epsilon}}{(2\pi)^{D-2\epsilon}}\; I_{D/2+1}[\mu^2]=-\frac{1}{(4\pi)^{D/2}}
\frac{1}{(D/2)!}
+O(\epsilon)\,.
\label{rationals}
\eqe 
Note that since the numerator of this integral depends on $\mu^2$, it is not directly detectable in $D$-dimensional unitarity cuts. Furthermore, such terms only arise from terms in the numerator with two or more powers of loop-momenta, which is indeed the case for $n_{(a)}$ in eq.(\ref{na}). Note that since rational terms only appear at one-loop in even-dimensions, this automatically leads to the conclusion that anomalies can only appear at one-loop in even-dimensions.

Straightforwardly carrying-out integral reduction in this manner, and isolating the rational terms, yields the anomalous, parity-odd, rational terms for $D=6,8$, and $10$:
\eqa\label{DAnoms}
\nonumber &&R^{anom}_{D = 6}= -\frac{\left[G_{(123)}+G_{(124)}+G_{(134)}\right]|_{k_1\rightarrow e_1} F_2\wedge F_3\wedge F_4}{3*4!*Gram_{(1,2,3)}}+(\rm{cyclic})\\
\nonumber &&R^{anom}_{D = 8}=-
\frac{\left[G_{(1234)}+G_{(1345)}+G_{(1235)}+G_{(1245)}\right]|_{k_1\rightarrow e_1}F_2\wedge F_3\wedge F_4\wedge F_5}{4*5!*Gram_{(1,2,3,4)}}+(\rm{cyclic})\\
\nonumber &&R^{anom}_{D = 10}=-\frac{\left[G_{(12345)}+G_{(13456)}+G_{(12356)}+G_{(12456)}+G_{(12346)}\right]|_{k_1\rightarrow e_1} F_2\wedge F_3\wedge \cdots \wedge F_6}{5*6!*Gram_{(1,2,3,4,5)}}\\ 
&&\,\,\,\,\,\,\,\,\,\,\,\,\,\,\,\,\,\,\,\,\,\,\,\,\,\,\,\,\,\,\,\,\,\,\,\,\,\,\,
+(\rm{cyclic}) \, ,
\eqae 
where $G_{(i_1,\cdots,i_l)}$ is the determinant of the Gram-matrix of $k_{i_1},\cdots ,k_{i_l}$, and the notation $G_{(12345)}|_{k_1\rightarrow e_1}$ indicates that we linearly replace $k_1$ with $k_1+a e_1$ in $G_{(12345)}$, and collect the coefficient in front of $a$. For example for $D=6$:
\eq
\left[G_{(123)}+G_{(124)}+G_{(134)}\right]|_{k_1\rightarrow e_1}=\left[tu(\e_1\cdot k_2)+st(\e_1\cdot k_3)+su(\e_1\cdot k_4)\right]\,.
\eqe
The presence of the Gram-determinant in the denominator is simply due to the integral reduction procedures. These rational terms are in fact invariant under permutations of the kinematic labels, and thus the rational term for the full color-dressed amplitude is given by:
\eq
\mathcal{R}^{anom}=R^{anom}str(T_1\cdots T_n)\, . \label{rProd}
\eqe
It is straightforward to verify that these terms precisely give the anomaly: the variation of the polarization vector, $\epsilon_i \rightarrow k_i$, automatically produces Gram determinants in the numerator that cancel against those in the denominator, leaving
\eq
\delta_{e_1\rightarrow k_1}\mathcal{R}^{anom}=F_2\wedge F_3\wedge \cdots \wedge F_n \times str(T_1\cdots T_n) \, . \label{rVary}
\eqe
Crucially if $str(T_1 \cdots T_n) = 0$, the original condition for anomaly cancellation, then the entire parity-odd rational term vanishes. Note that this is consistent with the previous on-shell analysis, which showed that if $str(T_1 \cdots T_4) = 0$, all spurious poles cancel and no rational terms are needed for locality.

\subsection{Gauge-invariant rational terms}

Anomaly cancellation is more subtle when $str(T_1\cdots T_n)$ does not vanish. Here, we construct the parity-odd gauge-invariant rational building-blocks seen in gauge-invariant expressions of parity-odd integral coefficients. Rather than construct gauge-invariant rational terms from unitarity cuts, we extract them in the more traditional way, from Feynman diagrams. 

As in the previous calculation, we begin with a study of anomaly cancellation in six-dimensions, where we make contact with both the Green-Schwarz mechanism, and with the previous on-shell results. Here, the anomalous rational term can be conveniently written as:
\eq
\mathcal{R}^{anom} = -\frac{1}{18}\bigg[\left(\frac{(\e_1\cdot k_2)}{s}+\frac{(\e_1\cdot k_3)}{u}+\frac{(\e_1\cdot k_4)}{t}\right)F_2\wedge F_3\wedge F_4+({\rm cyclic})\bigg] str(T_aT_bT_cT_d) \, . 
\label{Anomaly}
\eqe
If the symmetrized color-factor does not vanish, then the gauge-variation of the parity-odd rational term may be cancelled via the Green-Schwarz mechanism if and only if there exists a representation $t_a$ such that the trace factorizes $str(T_aT_bT_cT_d)=tr(t_at_b)tr(t_ct_d)$. As discussed above, the Green-Schwarz mechanism introduces a two-form with both parity-even and parity-odd couplings to gauges field via,
\eqa
{\cal L}_{\rm I} &=& H_{abc} H^{abc} = \left[dB + tr(A \wedge dA)\right]^2 \,\, , \, {\rm and} \,\, \nonumber\\
{\cal L}_{\rm II} &=& B \wedge tr(F^{(1)} \wedge \cdots \wedge F^{(D/2-1)}) \label{L12} \, , \,\, {\rm which \,\, for} \,\, D = 6 \,\, {\rm reduces \,\, to} \\
&&\longrightarrow B \wedge tr(F^{(1)} \wedge F^{(2)}) \, .
\eqae
As repeatedly stated above, these two interactions cannot be made mutually gauge-invariant, and their gauge-\emph{variance} cancels against that of the anomalous rational term.

To see how this happens, consider the parity-odd $s$-channel Feynman trees coupling gluons via two-form exchange. Parity violation mandates that both distinct interactions enter into the tree amplitude; the color-structure of the interactions dictates that these and only these parity-odd trees are proportional  $tr(t_1t_2)tr(t_3t_4)$. The detailed structure of the $s$-channel tree contribution is, 
\eq
GS_{(12)}^{\,6d \, {\rm tree}}=\frac{tr(t_1t_2)tr(t_3t_4)}{6 \, s}\bigg[(F_{1}\wedge F_{2}\wedge F_4)(\e_3\cdot k_{4})+(F_{1}\wedge F_{2} \wedge F_3)(\e_{4}\cdot k_3) + \big\{(1,2) \leftrightarrow (3,4)\big\} \bigg]\,.
\eqe
Adding $GS_{(12)}^{\,6d \, {\rm tree}}$ and $R^{anom}$, the term proportional to $tr(t_1t_2)tr(t_3t_4)$ is:
\eqa
\left(R^{anom}+GS_{(12)}^{\,6d \, {\rm tree}}\right) 
&=& \frac{1}{18}\left[\left(2\frac{(\e_1\cdot k_2)}{s}-\frac{(\e_1\cdot k_4)}{t}-\frac{(\e_1\cdot k_3)}{u}\right)F_2\wedge F_3\wedge F_4\right] \nonumber \\
&+& \frac{1}{18}\left[\left(2\frac{(\e_2\cdot k_1)}{s}-\frac{(\e_2\cdot k_3)}{t}-\frac{(\e_2\cdot k_4)}{u}\right)F_3\wedge F_4\wedge F_1\right]\nonumber \\
&+& \frac{1}{18}\left[\left(2\frac{(\e_3\cdot k_4)}{s}-\frac{(\e_3\cdot k_2)}{t}-\frac{(\e_3\cdot k_1)}{u}\right)F_4\wedge F_1\wedge F_2\right]\nonumber \\
&+& \frac{1}{18}\left[\left(2\frac{(\e_4\cdot k_3)}{s}-\frac{(\e_4\cdot k_1)}{t}-\frac{(\e_4\cdot k_2)}{u}\right)F_1\wedge F_2\wedge F_3\right] .
\eqae
It is amusing that the Green-Schwarz contribution simply tweaks the coefficient in the anomalous rational term such that the combination is gauge invariant. Converting the above into on-shell form, one finds:
\eq
R^{anom}+GS_{(12)}^{\, 6d \, {\rm tree}}=-\frac{(t-u)}{18stu}F^4\,. \label{6Dmatch}
\eqe
We see that the gauge invariant combination is precisely the rational term in eq.(\ref{FinalR})! Imposing locality on the result obtained from unitarity cuts inevitably leads to the rational term that is precisely the remnant combination of the anomalous rational term and the Green-Schwarz tree-amplitude. In other words, the amplitude not only ``knows" that there is a new particle in the spectrum, it also knows that there is a remnant from the cancelation and its precise form.

Cancellation between the parity-odd, anomalous, rational term and the Green-Schwarz mechanism is slightly more subtle in eight-dimensions and higher: as noted in the original literature, new terms in the action are needed~\cite{GS, Forms}. Recall that the full ${\cal R}^{anom}$ is given by $R^{anom}(1 \cdots n) str(T_1 \cdots T_n)$. For gauge-theories which admit Green-Schwarz anomaly cancelation, in factorized form, this can be rewritten as,
\eq
\mathcal{R}^{anom}=R^{anom}(1 \cdots n) \sum_{a b} tr(t_a t_b) str(t_{c_3} \cdots t_{c_{n}}) \, , \label{rFact}
\eqe
where the sum is over all distinct $n\choose2$ pairs of particle labels. Under a gauge-variation, $\epsilon_1 \rightarrow k_1$, the anomalous rational term becomes,
\eq
\mathcal{R}^{anom}\big|_{\epsilon_1 \rightarrow k_1} = \big( F_2 \wedge \cdots \wedge F_n \big) \times \sum_{a b} tr(t_a t_b) str(t_{c_3}\cdots t_{c_{n}}) \, .
\eqe
Thus we see that a gauge-variation of the anomalous rational terms will have \emph{all} possible double-trace structures.

However the gauge variation of any Green-Schwarz tree, say $\epsilon_1\rightarrow k_1$, must be to proportional to a particular subset of double-trace structures: $tr(t_1 t_j) str(t_{c_3}\cdots t_{c_n})$. In other words, the classic Green-Schwarz trees are insufficient to cancel all gauge-anomalies in eight-dimensions and higher! Alternatively, it is not possible for the coefficient of any particular double-trace structure, arising from the conventional Green-Schwarz tree and the anomalous rational term, to be made gauge-invariant for all possible variations.

Not surprisingly, the missing cancellation comes from local counter-terms which are free of poles:
\eqa
{\cal L}_{\rm III} &=& \omega_3^{\{12\}} \wedge \omega_{D-3}^{\{3...n\}} \label{L3} \\ 
&=& \bigg\{ \big[tr(A_1 \wedge dA_2) \wedge tr(A_3 \wedge F_4 \wedge \cdots \wedge F_n) + (1 \leftrightarrow 2)\big] + \mathcal{P}(3 \cdots n)\bigg\} \, , \nonumber
\eqae
Under gauge-variations, these interactions transform in the following way:
\eqa
\omega_3^{\{45\}} \wedge \omega_{5}^{\{123\}}\big|_{\epsilon_1 \rightarrow k_1} &=& -3 \big( F_2 \wedge F_3 \wedge F_4 \wedge F_5 \big) \times \sum_{b = 2}^{5} tr(t_4 t_5) str(t_{1} t_{2} t_{3})  \nonumber\,,\\
\omega_3^{\{45\}} \wedge \omega_{5}^{\{123\}}\big|_{\epsilon_5 \rightarrow k_5} &=& +2 \big( F_1 \wedge F_2 \wedge F_3 \wedge F_4 \big) \times \sum_{b = 2}^{5} tr(t_4 t_5) str(t_{1} t_{2} t_{3}) \, , \nonumber
\eqae
where, for concreteness, we have chosen a particular wedge-product in eight-dimensions. Using these new contact terms, we may now construct gauge-invariant rational terms in arbitrary even-dimensional spaces. The gauge-invariant coefficient of, for instance, the $tr(12) str(345)$ double-trace term present in the eight-dimensional five-point one-loop parity-odd rational term is:
\eqa
{\rm Inv}^{8d}_{(12)} &=& R^{\rm anom} - \bigg\{ \frac{5}{2} \big(GS^{\,8d \, {\rm tree}}_{(12)} \big) + \frac{1}{2}\big( \omega_3^{\{12\}} \wedge \omega_5^{\{345\}} \big) \bigg\} \,\, , \, {\rm where} \, \\
GS^{\,8d \, {\rm tree}}_{(12)} &=& \bigg(\frac{2 \epsilon_1 \cdot k_2}{s_{12}} F_2 \wedge F_3 \wedge F_4 \wedge F_5 + 
\frac{2 \epsilon_2 \cdot k_1}{s_{12}} F_3 \wedge F_4 \wedge F_5 \wedge F_1 \bigg) \, ,
\eqae
where these expressions come explicitly from the Feynman rules for the interaction terms in eqs.~\eqref{L12} and~\eqref{L3}. The corresponding structure which appears in ten dimensions is given by,
\eqa
{\rm Inv}^{10d}_{(12)} &=& R^{\rm anom} - \bigg\{ \frac{6}{2} \big(GS^{\,10d \, {\rm tree}}_{(12)} \big) + \frac{1}{2}\big( \omega_3^{\{12\}} \wedge \omega_7^{\{3456\}} \big) \bigg\} \,\, , \, {\rm where} \, \\
GS^{\,10d \, {\rm tree}}_{(12)} &=& \bigg( \frac{2 \epsilon_1 \cdot k_2}{s_{12}} F_2 \wedge F_3 \wedge F_4 \wedge F_5 \wedge F_6 + 
\frac{2 \epsilon_2 \cdot k_1}{s_{12}} F_3 \wedge F_4 \wedge F_5 \wedge F_6 \wedge F_1\bigg). \,\,\,\,
\eqae
One may explicitly check that these sums are manifestly gauge-invariant in all lines. 
\subsection{Integral coefficients from invariant rational terms}
These gauge-invariant parity-odd rational objects, ``remnants'', constructed explicitly from interference between the gauge-anomalies between Feynman diagrams, furnish a set of building-blocks for the gauge-invariant integral-coefficients. Intriguingly, the structure of the highest-oder integral coefficient (the box in $D = 6$) in a given cyclic ordering may be neatly rewritten,
\eq
C_4^{D = 6}(1,2,3,4) = \frac{s_{12}s_{23}s_{34}s_{41}}{Gram_{(1,2,3)}} \times ({\rm Inv}^{6d}_{(12)}+{\rm Inv}^{6d}_{(23)}+{\rm Inv}^{6d}_{(34)}+{\rm Inv}^{6d}_{(41)})
\eqe
where, as commented above, $Gram_{(1,2,3)} = s t u = s_{12} s_{14} s_{13}$, and the six-dimensional invariants may be rewritten in terms of the spinor-helicity variables, as reflected in eq.~\eqref{6Dmatch}. Motivated by this natural form for the parity-odd box-coefficient in six-dimensions, one would naturally conjecture that the pentagon-coefficient in eight-dimensions, etc., would be related to those invariant building-blocks in a similarly nice way. Indeed, the ansatz,
\eq
C_5^{D = 8}(1,2,3,4,5) = \frac{s_{12}s_{23}s_{34}s_{45}s_{51}}{Gram_{(1,2,3,4)}} \times ({\rm Inv}^{8d}_{(12)}+{\rm Inv}^{8d}_{(23)}+{\rm Inv}^{8d}_{(34)}+{\rm Inv}^{8d}_{(45)}+{\rm Inv}^{8d}_{(51)})
\eqe
exactly reproduces the parity-odd scalar pentagon-coefficient extracted from the $(1,2,3,4,5)$-pentagon \emph{diagram} in the chiral gauge-theory. In other words, the gauge-invariant integral coefficients contain explicit information about these ``remnant'' rational terms, leftover after anomaly cancellation between the Green-Schwarz trees and the chiral-fermion loops.

Natural extensions of this ansatz to the (famous) parity-odd hexagon-coefficient~\cite{GS} fail, due to the complicated singularity structure at six-points, inherent in the scalar hexagon-integral. However, as mentioned in the introduction, recent methods~\cite{HennDiff} exploit differential equations to capture the behavior of integrals near their singularities without the need to explicitly integrate them. It would be extremely interesting to use these methods to uniquely fix the parity-violating sectors in ten-dimensions. Successful implementation of this program would forcefully assert the conclusion that the highest-order integral-coefficient in the parity-odd sector of a theory entirely dictates the structure of the whole amplitude. We leave such explorations for future work, and move-on to address the pressing issue of how the perturbative gravitational anomalies, in theories with chiral- yet CPT-invariant spectra, manifest themselves in the on-shell S-matrix in, for example, six-dimensions~\cite{AlvarezGaume:1983ig}.

\section{$D=6$ Gravitational anomaly}
In this section we will be interested in gravity coupled to various chiral-matter in six-dimensions. We will compute the cut-constructible piece of the parity-odd one-loop four-point graviton amplitude, which will be unique to chiral-matter. In six-dimensions, there are three types of chiral coupling for gravity theories. Chiral-matter includes chiral-fermions as well as (anti) self-dual two-forms, while for supergravity theories there is also the presence of chiral-gravitinos. The presence of self-dual two-forms introduces an interesting question which can be readily addressed from our perspective: for a generic number of (anti) self-dual two-forms, there are no covariant actions and hence Feynman rules are unavailable. However, from an S-matrix point of view, there is no problem of defining consistent tree-level scattering amplitudes involving chiral two-forms, and this is all that is needed for the construction of it's quantum corrections. We will proceed by applying unitarity methods to compute the coefficients in the scalar-integral basis. This again defines the S-matrix up to rational ambiguities, which we will fix by enforcing locality.

It is well known that at the quantum level, chiral two-forms induce an anomaly when coupled to gravity. Thus a non-trivial test is to demonstrate that one recovers both the gravitational anomaly, and their attendant anomaly cancellation conditions, in the process of demanding unitarity and locality. As we now show, this is indeed the case.
\subsection{$D=6$ chiral tree-amplitude for $M_{4}(hhXX)$}
We begin by presenting the four-point amplitude for two gravitons coupled to a pair of chiral gravitinos, self-dual two-forms, and chiral spinors respectively:
\eqa
 &&M_4(h_{a_1 b_1 \dot{a}_1 \dot{b}_1},h_{a_2 b_2 \dot{a}_2 \dot{b}_2},\psi_{a_3 b_3 \dot{a}_3},\psi_{a_4 b_4 \dot{a}_4})
=\\
\nonumber&&\quad\quad\quad\quad\quad\quad\quad\quad  \frac{\langle 1_{a_1} 2_{a_2} 3_{a_3} 4_{a_4}\rangle 
\langle 1_{b_1} 2_{b_2} 3_{b_3} 4_{b_4}\rangle  
[1_{\dot{a}_1} 2_{\dot{a}_2} 3_{\dot{a}_3} 4_{\dot{a}_4}] 
[1_{\dot{b}_1}|3| 2_{\dot{b}_2} ]}{s \,\, t \,\, u}\bigg|_{sym} \label{6DgraviTino}\,,\\
&&M_4(h_{a_1 b_1 \dot{a}_1 \dot{b}_1},h_{a_2 b_2 \dot{a}_2 \dot{b}_2},B_{a_3b_3},B_{a_4b_4})
=\\\nonumber &&\quad\quad\quad\quad\quad\quad\quad\quad  
 \frac{
\langle 1_{a_1} 2_{a_2} 3_{a_3} 4_{a_4}\rangle 
\langle 1_{b_1} 2_{b_2} 3_{b_3} 4_{b_4}\rangle 
[1_{\dot{a}_1}|3| 2_{\dot{a}_2}] 
[1_{\dot{b}_1}|4| 2_{\dot{b}_2}]}{s \,\, t \,\, u}\bigg|_{sym} \label{6DgrTensor}\,,\\
&&M_4(h_{a_1 b_1 \dot{a}_1 \dot{b}_1},h_{a_2b_2 \dot{a}_2 \dot{b}_2},\chi_{a_3},\chi_{a_4})
=\\
\nonumber  &&\quad\quad\quad\quad\quad\quad\quad\quad  \frac{
\langle 1_{a_1} 2_{a_2} 3_{a_3} 4_{a_4}\rangle 
\langle 1_{b_1} |3|2_{b_2}\rangle 
[1_{\dot{a}_1}|4| 2_{\dot{a}_2}] 
[1_{\dot{b}_1}|4| 2_{\dot{b}_2} ]}{s \,\, t \,\, u}\bigg|_{sym} \label{6DgrFermion}\,,
\eqae
where $|_{sym}$ indicates the symmetrization of the SU(2) indices if more than one is present for a given leg. The symmetrization reflects the fact that the graviton, gravitino, self-dual two-form and chiral spinor transforms as $(3,3)$, $(2,3)$, $(0,3)$ and $(0,2)$  under the SU(2)$\times$SU(2) little group.  

The validity of the above amplitude can be confirmed by showing that on all three factorization poles, it factorizes correctly into a product of three-point amplitudes involving two chiral states coupled to a graviton. The three-point amplitudes are uniquely determined from little group requirements~\cite{YtBartek}:
\eqa
M_3(h_{a_1 b_1 \dot{a}_1 \dot{b}_1},\psi_{a_2 b_2 \dot{a}_2},\psi_{a_3 b_3 \dot{a}_3})
&=& (\Delta_{a_1 a_2 a_3} \tilde{\Delta}_{\dot{a}_1 \dot{a}_2 \dot{a}_3})(\Delta_{b_1b_2b_3}\tilde{u}_{1 \dot{b}_1})\bigg|_{sym}\,, \\
M_3(h_{a_1 b_1 \dot{a}_1 \dot{b}_1},B_{a_2b_2},B_{a_3b_3 })
&=& (\Delta_{a_1 a_2 a_3} \tilde{u}_{1 \dot{a}_1})(\Delta_{b_1b_2b_3}  \tilde{u}_{1 \dot{b}_1})\bigg|_{sym}\,,\\
M_3(h_{a_1 b_1 \dot{a}_1 \dot{b}_1},\chi_{a_2},\chi_{a_3})
&=& (\Delta_{a_1 a_2 a_3} \tilde{u}_{1 \dot{a}_1})(u_{1 b_1} \tilde{u}_{1 \dot{b}_1})\bigg|_{sym}\,.
\eqae
Alternatively, the above four-point amplitudes can be derived from the fully supersymmetric $\mathcal{N}=(2,2)$ supergravity amplitude. In superspace, for the sake of being explicit, the full multiplet is given by an expansion of $\eta^{ia},\tilde{\eta}^{\hat{i}\dot{a}}$s, where $i,\hat{i}=1,2$. Thus we have in total 8 grassmann variables, and the superfield is expanded as 
\eqa
\nonumber&& \Phi(\eta^{ia},\tilde{\eta}^{\hat{i}\dot{a}})=\phi+\eta^{ia}\psi_{ia}+\tilde{\eta}^{\hat{i}\dot{a}}\tilde{\psi}_{\hat{i}\dot{a}}+\eta^{2(ij)}\phi_{(ij)}+\tilde{\eta}^{2(\hat{i}\hat{j})}\tilde\phi_{(\hat{i}\hat{j})}+\eta^{i(a}\eta^{b)}_{i}B_{(ab)}+\tilde{\eta}^{\hat{i}(\dot{a}}\tilde\eta^{\dot{b})}_{\hat{i}}\tilde\phi_{(\dot{a}\dot{b})}+\tilde{\eta}^{\dot{a}}_{\hat{i}}\eta^{a}_jA^{\hat{i}j}_{a\dot{a}}\\
\nonumber&+&\eta^{2(ij)}\tilde{\eta}^{\dot{a}\hat{i}}\tilde\psi_{(ij)\hat{i}\dot{a}}+\tilde{\eta}^{2(\hat{i}\hat{j})}\eta^{ai}\psi_{(\hat{i}\hat{j})ia}+\eta^{ai}\eta_{a}^{j}\eta^{b}_i\psi_{jb}+\tilde\eta^{\dot{a}\hat{i}}\tilde\eta^{\hat{j}}_{\dot{a}}\tilde\eta_{\hat{i}}^{\dot{b}}\tilde\psi_{\hat{j}\dot{b}}+\eta^{2(ab)}\tilde{\eta}^{\hat{j}\dot{a}}\psi_{(ab)\dot{a}\hat{j}}+\tilde{\eta}^{2(\dot{a}\dot{b})}\eta^{ia}\tilde\psi_{i(\dot{a}\dot{b})a}\\
\nonumber&+&\eta^{2(ij)}\tilde{\eta}^{2(\hat{i}\hat{j})}\phi_{(ij)(\hat{i}\hat{j})}+\eta^{2(ab)}\tilde{\eta}^{2(\dot{a}\dot{b})}g_{(ab)(\dot{a}\dot{b})}+\eta^{2(ab)}\tilde{\eta}^{2(\hat{i}\hat{j})}B_{(ab)(\hat{i}\hat{j})}+\eta^{2(ij)}\tilde{\eta}^{2(\dot{a}\dot{b})}B_{(ij)(\dot{a}\dot{b})}\\
\nonumber&+&\eta^{ia}\eta^{j}_{a}\eta_i^{b}\tilde\eta^{\hat{i}\dot{a}}A_{\hat{i}jb\dot{a}}+\tilde\eta^{\hat{i}\dot{a}}\eta^{\hat{j}}_{\dot{a}}\eta_{\hat{i}}^{\dot{b}}\eta^{ia}A_{\hat{j}ia\dot{b}}+\eta^4\phi'+\tilde{\eta}^4\tilde{\phi}'+\cdots\,,
\eqae  
where $\cdots$ are degree 5 and higher in grassmann variables. One can see the bosonic states are $2\times1+2\times(4\times3+4\times4)+(4\times3\times3+2\times4\times4)=126$, as expected for maximal super gravity. The $\mathcal{N}=(2,2)$ supergravity four-point amplitude is given by
\eq
\mathcal{A}_4=\frac{\delta^{8}(Q)\delta^{8}(\tilde{Q})}{stu},\quad\quad\delta^{8}(Q)\equiv \prod_{i=1}^2\prod_{A=1}^4\left(\sum_{n=1}^4\lambda_n^{Aa}\eta_{na}^{i}\right),\quad \delta^{8}(\tilde{Q})\equiv \prod_{\hat{i}=1}^2\prod_{A=1}^4\left(\sum_{n=1}^4\tilde\lambda_{nA}^{\dot{a}}\tilde\eta_{n\dot{a}}^{\hat{i}}\right)\,.
\eqe
The component amplitudes can be obtained as:
\beqa
\nonumber eq.(\ref{6DgraviTino})&=&\mathcal{A}_4|_{\eta_1^{2(ab)}\tilde{\eta}_1^{2(\dot{a}\dot{b})}\eta_2^{2(ab)}\tilde{\eta}_2^{2(\dot{a}\dot{b})}\eta_3^{2(ab)}\tilde{\eta}_3^{\hat{j}\dot{a}}\tilde{\eta}_{3\dot{a}}^{\hat{i}}\tilde{\eta}_{3\hat{j}}^{\dot{b}} \eta_4^{2(ab)}\tilde{\eta}_4^{\hat{j}\dot{a}} }\,,\\
\nonumber eq.(\ref{6DgrTensor})&=&\mathcal{A}_4|_{\eta_1^{2(ab)}\tilde{\eta}_1^{2(\dot{a}\dot{b})}\eta_2^{2(ab)}\tilde{\eta}_2^{2(\dot{a}\dot{b})}\eta_3^{2(ab)}\tilde{\eta}_3^{2(\hat{i}\hat{j})}  \eta^{2(ab)}_4\tilde{\eta}_4^{2(\hat{i}\hat{j})} }\,,\\
\nonumber eq.(\ref{6DgrFermion})&=&\mathcal{A}_4|_{\eta_1^{2(ab)}\tilde{\eta}_1^{2(\dot{a}\dot{b})}\eta_2^{2(ab)}\tilde{\eta}_2^{2(\dot{a}\dot{b})}\tilde{\eta}_3^{2(\hat{i}\hat{j})}\eta_3^{ai}  \tilde{\eta}_4^{2(\hat{i}\hat{j})}\eta_4^{ai}\eta_{4a}^{j}\eta_{4i}^{b}} \,.
\eeqa
We now proceed to use these results to construct the parity-odd four-point one-loop amplitude, where the potential anomalies reside.

\subsection{Chiral loop-amplitudes }
Using the four-point tree-amplitudes with two graviton and two chiral-matter, we can now proceed and compute the one-loop four-graviton amplitude with chiral-matter in the loop. Again we begin by expanding the integrals in the scalar integral basis:
\eqa\label{GravityLoopAmp}
\nonumber \mathcal{M}^{1-{\rm L, odd};X}_{4}&=&C_4^{X}(1,2,3,4)I_{1,2,3,4}+C_4^{X}(1,3,4,2)I_{1,3,4,2}+C_4^{X}(1,4,2,3)I_{1,4,2,3}\\
&+&C_{3s}^{X}I_{3s}+C_{3u}^{X}I_{3u}+C_{3t}^{X}I_{3t}+C_{2s}^{X}I_{2s}+C_{2u}^{X}I_{2u}+C_{2t}^{X}I_{2t}+R\,,
\eqae
where $X=(\psi, B, \chi)$ represents the contribution from the chiral gravitino, two-form, and fermion respectively. Again to simplify our task, we single out a four-dimensional sub plane spanned by the momenta of the external legs. This allows us to project the six-dimensional states onto four-dimensional ones: the $3\times3=9$-degrees of freedom then becomes the graviton, two vectors, and three-scalars respectively:
\eq 
(h_{11,\dot{1}\dot{1}}\,,\; h_{22,\dot{2}\dot{2}})\,\,,\; (h_{12,\dot{2}\dot{2}}\,,\; h_{12,\dot{1}\dot{1}})\,\,,\; ( h_{11,\dot{1}\dot{2}}\,,\; h_{22,\dot{1}\dot{2}})\,\,,\;  (h_{12,\dot{1}\dot{2}}\,,\; h_{11,\dot{2}\dot{2}}\,,\; h_{22,\dot{1}\dot{1}})\,.
\eqe
Since we are considering the parity-odd part of the amplitude, to ensure that we have enough vectors to span all six directions, we will choose all four six-dimensional graviton states to be distinct. We begin with the following helicity configuration:
$$(1_{12\dot 1\dot 2}\,,2_{22\dot 1\dot 1}\,,3_{22\dot 2\dot 2}\,, 4_{1 1\dot 2\dot 2})\,.$$
The computation of the integral coefficients is straightforward and similar to that of the QCD and QED computation. The slight distinction is that the highest power of $m\tilde{m}$ appearing in the unitarity cuts is three times that of the gauge theories, which leads to the result that after integral reduction, the power of poles for the scalar integrals will be higher. As an example, the $m\tilde{m}$-dependent integral coefficients for the self-dual two-form are given as:
\beqa
\notag &&\tilde C_4^B(1,2,3,4) =-\frac{4 (m\tilde m)^2 (s -t) (s t + 2  um\tilde m)R^{(4)}}{s t u^2}\, ,\\
\notag &&\tilde C_4^B(1,3,4,2) =-\frac{4 (m\tilde m)^2 [s^2 u + 2  t ( s-t)m\tilde m]R^{(4)}}{s t^2 u}\,,\\
\notag &&\tilde C_4^B(1,3,2,4) =\frac{4 (m\tilde m)^2 [t^2 u - 2 s (s-t)m\tilde m ]R^{(4)}}{s^2 t u}\,,\\
\notag &&\tilde C_{3s}^B=-\frac{4 (m\tilde m)^2 (s^2 + 3 t^2 + u^2)R^{(4) }}{t^2 u^2}\,,~~\tilde C_{3t}^B = \frac{4 (m\tilde m)^2 (3 s^2 + t^2 + u^2)R^{(4)}}{s^2 u^2}\,,\\
&&\tilde C_{3u}^B = -\frac{8 (m\tilde m)^2 u (s-t)R^{(4)} }{s^2 t^2}\,,
\eeqa
where $R^{(4)}=\langle 13\rangle^2 \langle 23\rangle^2 [12]^2$.  
In this special case, the coefficients of different loop species have a simple relation:
\beqa
 5 \tilde C^B=-20\tilde C^\chi=-4\tilde C^\psi\,.
\eeqa
After integral reduction, the $m\tilde{m}$-dependent integral coefficients now becomes:
\beqa
\notag&& C_{2s}^B = -\frac{2s (12 s^4 + 53 s^3 t + 91 s^2 t^2 + 75 s t^3 + 49 t^4) R^{(4)}}{105 t^3 u^3}\,,\\
\notag&& C_{2t}^B = \frac{2t (49 s^4 + 75 s^3 t + 91 s^2 t^2 + 53 s t^3 + 12 t^4)R^{(4)}}{105 s^3 u^3}\,,\\
\notag && C_{2u}^B = -\frac{2u^2(12 s^3 - 11 s^2 t + 11 s t^2 - 12 t^3) R^{(4)}}{105 s^3 t^3}\,,\\
\notag && C_{3s}^B =-\frac{4s^2 (4 s^6 + 23 s^5 t + 55 s^4 t^2 + 70 s^3 t^3 + 50 s^2 t^4 + 
    15 s t^5 + 7 t^6)R^{(4)}}{105 t^4 u^4}\,,\\
\notag&& C_{3t}^B = \frac{4t^2 (7 s^6 + 15 s^5 t + 50 s^4 t^2 + 70 s^3 t^3 + 55 s^2 t^4 + 
   23 s t^5 + 4 t^6)R^{(4)}}{105 s^4 u^4}\,,\\
\notag&& C_{3u}^B = -\frac{4u^3(4 s^5 - s^4 t + s^3 t^2 - s^2 t^3 + s t^4 - 4 t^5) R^{(4)}}{ 105 s^4 t^4}\,,\\
 \notag && C_4^B(1,2,3,4) =-\frac{8 s^2 t^2 (s - t)R^{(4)}}{105 u^4}\, ,~~ C_4^B(1,3,4,2) =-\frac{2s^2u^2 (4 s + 3 t)R^{(4)} }{105 t^4}\,,\\
 && C_4^B(1,3,2,4) =\frac{2t^2u^2 (3 s + 4 t) R^{(4)}}{105 s^4}\,,~~~~~5C^B=-20C^\chi=-4C^\psi\,.
\eeqa
Similarly, we can also computed the configuration $(1_{11\dot 2\dot 2}\,,2_{22\dot 1\dot 1}\,,3_{22\dot 2\dot 2}\,, 4_{1 1\dot 1\dot 1})\,$ and the
parity-odd integral coefficients of self-dual 2-form loops are
\beqa
\notag &&\tilde C_4^B(1,2,3,4) = 
 - \frac{R^{(4)}}{ t^2 u^4}[
     8 u^2 (t- u)(m\tilde m)^3  +
     2  s u (8 s^2 + 17 s u + 8 u^2)(m\tilde m)^2\\
  \notag    &&~~~~~~~~~~~~~~~~~~~~~~~~~~~~~~~~~~~~~~~~~~~~~~~~~~~~~~~~~~~~~ + 4  s t^3 (s - u)m\tilde m-s^2 t^4] \,,\\
\notag &&\tilde C_4^B(1,3,4,2) = -\frac{R^{(4)}}{ t^4 u^2}[8 t^2 ( t-u)(m\tilde m)^3  - 
     2  s t (8 s^2 + 17 s t + 8 t^2)(m\tilde m)^2\\
\notag    &&~~~~~~~~~~~~~~~~~~~~~~~~~~~~~~~~~~~~~~~~~~~~~~~~~~~~~~~~~~~~~  - 4  s u^3(s - t) m\tilde m + 
     s^2 u^4]\,,\\
\notag &&\tilde C_4^B(1,3,2,4) =-\frac{2(t-u) (s+4 m\tilde m)(m\tilde m)^2R^{(4)}}{t^2 u^2}\,,\\
\notag &&\tilde C_{3s}^B =-\frac{2R^{(4)}}{t^4 u^4}[4  t  u(t - u) (2 t^2 + 3 t u + 2 u^2)(m\tilde m)^2\\
\notag&&~~~~~~~~~~~~ +
  2 s (t - u) (2 t^4 + 5 t^3 u + 5 t^2 u^2 + 5 t u^3 + 2 u^4)m\tilde m  +
  s^2 (t^5 - u^5)]\,,\\
\notag&&\tilde C_{3t}^B = \frac{2[4  u (2 t -  u)(m\tilde m)^2 + 2  t^2 (2s- u)m\tilde m -s t^3]R^{(4)}}{
     t u^4}\,,\\
\notag&&\tilde C_{3u}^B = \frac{2[4  t (t - 2 u)(m\tilde m)^2- 2  u^2 (2s -t)m\tilde m + s u^3]R^{(4)} }{
     t^4 u}\,,\\
\notag&&\tilde C_{2s}^B = -\frac{(t - u) [s (2 t^2 + t u + 2 u^2) + 
       4  (2 t^2 + 3 t u + 2 u^2)m\tilde m]R^{(4)}}{ t^3 u^3}\,,\\
&&\tilde C_{2t}^B = -\frac{(s-t)(4 m\tilde m - t) R^{(4)}}{t u^3}\,,~~\tilde C_{2u}^B = \frac{ (s-u)(4 m\tilde m - u)R^{(4)}}{t^3 u}\,,
\eeqa
where $R^{(4)}=\langle 1 3\rangle^4[14]^2$. We list the contributions from chiral-fermion and chiral graviton in appendix~\ref{RIC}. After the integal reduction of $m\tilde m$ terms, the integral coefficients of self-dual 2-form become
\beqa\label{BCoeff}
\notag &&C_4^{B}(1,2,3,4)=-\frac{s^2 (20 t^3 - 14 t^2 u + 8 t u^2 + 7 u^3)R^{(4)}}{105 u^5}\,,\\
\notag &&C_4^{B}(1,3,4,2)=\frac{s^2 (7 t^3 + 8 t^2 u - 14 t u^2 + 20 u^3)R^{(4)}}{105 t^5}\,,\\
\notag &&C_4^{B}(1,4,2,3)=-\frac{(7 t^3 + t^2 u - t u^2 - 7 u^3)R^{(4)}}{105 s^3}\,,\\
\notag &&C_{3s}^{B}=\frac{2 (20 t^9 + 26 t^8 u + 9 t^6 u^3 + 15 t^5 u^4 - 15 t^4 u^5 - 
   9 t^3 u^6 - 26 t u^8 - 20 u^9)R^{(4)}}{105 t^5 u^5}\,,\\
 \notag  &&C_{3t}^{B}=\frac{2 t (20 t^6 + 66 t^5 u + 72 t^4 u^2 + 35 t^3 u^3 + 47 t^2 u^4 + 
   61 t u^5 + 35 u^6)R^{(4)}}{105 s^3 u^5}\,,\\
 \notag  &&C_{3u}^{B}=-\frac{2 u (35 t^6 + 61 t^5 u + 47 t^4 u^2 + 35 t^3 u^3 + 72 t^2 u^4 + 
    66 t u^5 + 20 u^6)R^{(4)}}{105 s^3 t^5}\,,\\
\notag &&C_{2s}^{B}=-\frac{2(30 t^6 -11 t^5 u +11 t u^5 -30 u^6)R^{(4)}}{105 t^4 u^4}\,,\\
\notag   &&C_{2t}^{B}=\frac{2 (30 t^4 + 59 t^3 u + 21 t^2 u^2 - 12 t u^3 + 14 u^4)R^{(4)}}{105 s^2 u^4}\,,\\
 &&C_{2u}^{B}=  -\frac{2 (14 t^4 - 12 t^3 u + 21 t^2 u^2 + 59 t u^3 + 30 u^4)R^{(4)}}{105 s^2 t^4}\,.
 \eeqa
The contributions for chiral-fermion and chiral graviton are given in appendix~\ref{RIC}.
Thus we have the parity-odd contribution of the chiral-two from, up to the rational term in eq.(\ref{GravityLoopAmp}). We will now obtain it using locality.

\subsection{The rational term and gravitational anomaly}
We will now fix our rational term by enforcing locality on the cut constructed piece of the amplitude. However, before doing so, it is well known that gravity theories in $4k+2$-dimensions can develop gravitational anomalies when coupled to chiral~\cite{AlvarezGaume:1983ig}. Recall, once more, that the fingerprint of anomalies in on-shell amplitudes is the tension between enforcing locality and unitarity on the full amplitude. We must see the presence of anomalies in our attempt to construct the rational terms. Let us consider the various factorization limit at four-points (non-singular terms are dropped). The factorization limit of $(1_{12\dot 1\dot 2}\,,2_{22\dot 1\dot 1}\,,3_{22\dot 2\dot 2}\,, 4_{1 1\dot 2\dot 2})\,$ is
 \beqa\label{GravyPoles} 
 \nonumber  M_{4,\,Cut}^{1-{\rm L, odd},B}\Big|_{s\rightarrow 0} &=&-\frac{u^2 (5 s + 4 u)R^{(4)}}{315 s^3}\,,\\
 \nonumber M_{4,\,Cut}^{1-{\rm L, odd},B}\Big|_{u\rightarrow 0} &=&\frac{2 t (4 t^2 + 6 t u + u^2)R^{(4)}}{315 u^3}\,,\\
  M_{4,\,Cut}^{1-{\rm L, odd},B}\Big|_{t\rightarrow 0} &=&\frac{u^2 (5 t + 4 u)R^{(4)}}{315 t^3}\,.
  \eeqa

The presence of higher-order poles is a clear violation of locality, and needs to be rectified by additional rational terms. However, with a little thought, it is simple to see that there are no gauge invariant rational terms to cancel the degree-3 poles. To see this, it is worthwhile to note that the parity-odd gauge invariant rational terms for Yang-Mills theory constructed so far are nicely built from the following gauge invariant combination:
\eq
F_i\wedge F_j \wedge F_k\,,\quad  (\epsilon_i\cdot k_j-\epsilon_j\cdot k_i)\,, \quad (s_{ij}\;\epsilon_i\cdot \epsilon_j-2(k_i\cdot \epsilon_j)(k_j\cdot \epsilon_i) )\,,
\eqe
where the last term is added for completeness of the basis. At linear level, we can factorize the gravity polarization tensor as $\epsilon_{\mu\nu}\rightarrow \epsilon_\mu\epsilon_\nu$, and thus an acceptable parity-odd rational term must be given by a sum of terms with single factor of $F_i\wedge F_j \wedge F_k$ multiplied with other invariants. Finally, assuming that the loop-amplitude is given by a covariant integrand, through integral reduction, one can deduce that the highest power of poles for the rational term must be at most one-less than that of the scalar-integral coefficients. Now since the integral coefficients have at most fourth-order poles, the rational function must have at most third-order poles. Furthermore, since in the four-dimensional representation, we have three-scalars and a graviton, the only terms involving polarization vectors that are non-vanishing when they are projected into scalar states are $F_i\wedge F_j \wedge F_k$ and $\epsilon_j\cdot\epsilon_i$. Since there can only be one factor of the former, for a non-vanishing contribution one must include at least one power of $(s_{ij}\;\epsilon_i\cdot \epsilon_j-2(k_i\cdot \epsilon_j)(k_j\cdot \epsilon_i) )$, with only the first term in the parenthesis contributing if $i$ and $j$ are scalars. However, the presence of an additional factor $s_{ij}$ means that the power of poles are further surpressed. Thus in summary we conclude that for the parity-odd gravity invariant rational term for helicity $(1_{12\dot 1\dot 2}\,,2_{22\dot 1\dot 1}\,,3_{22\dot 2\dot 2}\,, 4_{1 1\dot 2\dot 2})\,$:

\noindent\textit{Any\; factorization\; channel\; that\; is\; associated \;with \;two \; four-dimensional\; scalars\; on \;one-side \;can \;have \;at\; most \;degree-2 poles.}

Thus there is no way for us to cancel the degree-3 poles in eq.(\ref{GravyPoles})! In other words, they have to cancel amongst themselves. Not surprisingly, the known anomaly-free combination:
 \eq\label{AnomFree}
 21 M^\chi-M^\psi+4 M^B
 \eqe
precisely achieves this! Note that the projection on to four-dimensional space, while simplifies the computation, loses the information of the full six-dimensional covariance. Thus in this simplified computation, we don't expect to obtain the full six-dimensional anomaly conditions, but we should see the solution to the anomaly conditions ensuring locality of the final result, and indeed it does. In fact for the current case, the anomaly-free combination cancels all poles! 
 
We can consider the helicity configuration of of $(1_{11\dot 2\dot 2}\,, 2_{22\dot 1\dot 1}\,, 3_{22\dot 2\dot 2}\, , 4_{1 1\dot 1\dot 1})\,$:
   \beqa
 \nonumber M_{4,\,Cut}^{1-{\rm L, odd}\psi}\Big|_{s\rightarrow 0} &=&\frac{s+2t}{84 s^2} R^{(4)}\,,\\
 \nonumber  M_{4,\,Cut}^{1-{\rm L, odd}\psi}\Big|_{u\rightarrow 0} &=&\frac{(100 t^3 - 96 t^2 u - 181 t u^2 - 471 u^3)R^{(4)}}{2520 u^4}\,,\\
 \nonumber  M_{4,\,Cut}^{1-{\rm L, odd}\psi}\Big|_{t\rightarrow 0} &=&\frac{(471 t^3 + 181 t^2 u + 96 t u^2 - 100 u^3)R^{(4)}}{2520 t^4}\,,\\
 \nonumber  M_{4,\,Cut}^{1-{\rm L, odd},B}\Big|_{s\rightarrow 0} &=&-\frac{s+2t}{105 s^2}R^{(4)}\,,\\
 \nonumber M_{4,\,Cut}^{1-{\rm L, odd},B}\Big|_{u\rightarrow 0} &=&\frac{(93 t^3 - 22 t^2 u + 12 t u^2 + 40 u^3)R^{(4)}}{1260 t^4}\,,\\
 \nonumber M_{4,\,Cut}^{1-{\rm L, odd},B}\Big|_{t\rightarrow 0} &=&-\frac{(40 t^3 + 12 t^2 u - 22 t u^2 + 93 u^3)R^{(4)}}{1260 u^4}\,,\\ 
 \nonumber M_{4,\,Cut}^{1-{\rm L, odd},\chi}\Big|_{s\rightarrow 0} &=&\frac{s+2t}{420 s^2}R^{(4)}\,, \\
 \nonumber M_{4,\,Cut}^{1-{\rm L, odd},\chi}\Big|_{u\rightarrow 0} &=&\frac{(20 t^3 + 48 t^2 u + 31 t u^2 + u^3)R^{(4)}}{2520 u^4}\,,\\
  M_{4,\,Cut}^{1-{\rm L, odd},\chi}\Big|_{t\rightarrow 0} &=&-\frac{(t^3 + 31 t^2 u + 48 t u^2 + 20 u^3)R^{(4)}}{2520 t^4}\,.
 \eeqa
Again the anomaly-free combination in eq.(\ref{AnomFree}) precisely cancels the degree-4 poles. The remaining degree-3 and lower poles can be completely canceled by the following invariant rational term:
\beqa 
R&=&\frac{-4 t^5 - 4 t^4 u + t^3 u^2 - t^2 u^3 + 4 t u^4 + 4 u^5}{10 t^3 u^3}R^{(4)}\,.
\eeqa
Thus with the above rational term, we've obtained a unitary and local parity-odd one-loop four-point amplitude. Of crucial importance is the fact that the same cancellation occurred in distinct helicity configuration, thus compensating the fact six-dimensional covariance was not manifest by identifying a particular four-dimensional sub-plane.

\section{Conclusions}
In this paper, we have constructed the parity-odd one-loop amplitudes for chiral-gauge and gravity theories using unitarity-methods, which constructs the integrand entirely from on-shell building blocks that are ignorant of any notion of gauge symmetry. We show that the constraints of unitarity and locality necessarily introduce new factorization channels in the one-loop amplitude. This only occurs in the parity-odd sector of the amplitude, which would be absent for non-chiral theories. Consistency of the theory then relies on whether or not this new channel can be interpreted as an exchange of physical fields. For gauge theories, the absence of such new factorization channel imposes constraints on color factors that are exactly that of anomaly cancellation, the vanishing of the symmetric trace with $D/2+1$ generators. For group theory factors that do not satisfy this constraint, the new factorization channel is here to stay. In four-dimensions, we show that the residue of this new factorization pole cannot correspond to the exchange of physical fields, and thus the theory is inconsistent. In six-dimensions, the residue is exactly the exchange of a two-form between gluons (photons in QED). Thus if the symmetrized trace factorizes, then consistency is achieved and the new factorization channel automatically reveal a new particle in the spectrum, the Green-Schwarz two-form.  

This tension between unitarity and locality is in fact isolated in a unique part of the amplitude, the cut-free rational terms. Indeed whilst unitarity method automatically gives an S-matrix that respects unitarity, there are potential violations of locality, and it is the job of the rational terms to restore locality. The new feature in chiral theories is that the requisite rational terms come with their own baggage. More precisely, they hold locality ransom: if we want locality back, the price to pay is to enlarge our spectrum. In the traditional Feynman diagram approach, such rational terms can be identified with the combination of the anomalous rational term from the one-loop Feynman diagrams and the tree-contribution of the Green-Schwarz mechanism, leaving behind a gauge invariant remnant. This remnant is an afterthought in the traditional literature, however from the S-matrix point of view, it is of upmost importance and anomaly cancellation is simply an ``off-shell" way of producing the requisite rational term. An interesting question is whether Feynman-diagrams are the unique ``off-shell" way of obtaining them. We will come back to this shortly. 

Finally we also extend our analysis to gravity coupled to chiral-matter. We study the parity-odd one-loop four-graviton amplitude with chiral fermions, two-forms and gravitinos in the loop. The case of chiral two-form is very interesting as no covariant action exists for gravity coupled to arbitrary number of these fields. It is thus interesting to see if unitarity methods can be employed to construct the correct loop-level amplitudes for such theories. A non-trivial test would be if the constraint of locality for the one-loop amplitude, derived though unitarity methods, exactly corresponds to that of gravitational anomaly cancellation. Indeed we demonstrate that the parity-odd four-graviton amplitude, which is again unique for chiral-matter, contains spurious singularities that cannot be canceled by any invariant rational terms. Thus for consistency, such singularities must cancel between different chiral-matter loops. We have demonstrated that the known anomaly free combination indeed achieves this cancellation.

There are many interesting extensions. In the analysis of chiral-gauge theory, the absence of UV-divergences and spurious poles strongly constrain the scalar integral coefficients. In particular in $D=2k$ $(k>2)$ dimensions, the parity-odd piece of the $(k+1)$-point amplitude can be expanded on the basis of scalar $(k+1)$-, $k$- and $(k-1)$-gon integrals. The $k$- and $(k-1)$-gon integral have ultraviolet divergences: their coefficients must be such that the overall divergence cancels. The $(k+1)$-gon integral has spurious singularities, which, when combined with that of the two lower-gons, must be such that it cancels with that of the gauge invariant rational term that is completely determined by eq.(\ref{DAnoms}). Indeed, it can be shown that using such constraint, one can fully determined the six-dimensional parity odd amplitude. Recently, it has been shown that one can utilize differential equations~\cite{HennDiff} to capture the behavior of integrals near singularities without explicitly integrating the full answer. Thus it is perceivable that the interrelation of the integral coefficients can be simplified using such differential equations. 

The  Adler-Bardeen theorem states that anomalies are one-loop exact~\cite{AdlerBardeen}. In our language this is tantamount to the statement that only at one-loop can the rational terms introduce new singularities that violate locality. In our current study this fact is completely not transparent.  In $D=6$ dimensions, we have shown that the first non-trivial gauge invariant rational term factorizes completely into known three-point amplitudes. It should then be possible to construct a recursion relation for the rational terms. For supersymmetric theories, it is also possible to apply recursion relations for the planar loop-amplitude~\cite{Simon, NimaBCFW, Rutgers}. Through the lens of these two recursion relations, it will be extremely appealing to see how Adler-Bardeen theorem arises for supersymmetric chiral-Yang-Mills theories, i.e. $\mathcal{N}=(1,0)$  in six-dimensions and maximal SYM in ten-dimensions. 

In both unitarity methods and Feynman diagrams, the parity-odd gauge invariant rational term is obtained by adding extra terms to the amplitude. For the former, it is an extra term that is required for locality, while for the latter, it is obtained by adding an extra contribution from the Green-Schwarz mechanism. One might ask if there is any way in which one directly obtains the full gauge invariant amplitude in one go. One possibility is applying color-kinematic duality to chiral gauge and gravity theories. It would be interesting to see if the duality satisfying numerator, satisfying certain constrains, automatically gives the invariant rational term~\cite{HenrikYT}. 

Finally it is well known that gauge-anomalies are related to global anomalies. The presence of global anomalies in supergravity amplitudes has been recently discussed in detail for $\mathcal{N}=4$ supergravity~\cite{RaduN=4}. Again here the symmetry violating amplitude appears as rational functions. It has been proposed that such anomaly controls the structure of the UV-divergence of $\mathcal{N}=4$ supergraivty~\cite{4Loop}, and it is conceivable that the effect of such global anomalies in multi-loop amplitudes, is similar to the effect of gauge anomalies, which is governed by the Adler-Bardeen theorem. Thus it will be interesting to establish the connection at the level of the amplitude.

\section*{Acknowledgements}
It is a pleasure to thank Nima Arkani-Hamed, Lance Dixon, Johannes Henn and Henrik Johansson for discussions. Y H. is supported by the Department of Energy under contract DE-SC0009988. W-M. C. thanks  National Science Council and NCTS of Taiwan and NTU CTS for support. 
\appendix
\section{Coefficient extraction}\label{CE}
\begin{figure}[h]
\centering
\includegraphics[scale=0.7]{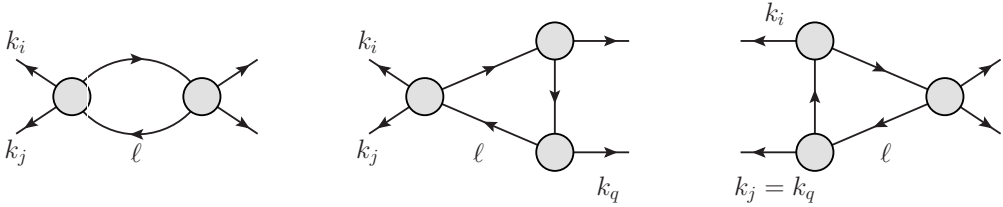}
\caption{Possible integrals contribute to bubble coefficients, here shows a specific ordering. For abelian gauge theories, one needs to consider permutation of the external legs $k_i$ and $k_j$ for the triangle in the right as well as $k_q$ and the unlabelled external leg for the triangle in the middle.  }\label{2-cut}
\end{figure}

In this paper, we have represented four-point one-loop massless scattering amplitude using scalar bubble, traingle and box integral as our basis. Here we give a brief review of how to extract the integral coefficients of these integral basis. We follow the work of Forde~\cite{FordeRules}, implemented in six-dimensions~\cite{YT6D, Scott6D}. To simplify the analysis, we will define a four-dimensional subspace that includes the four-external legs, such that the six-dimensional $2\times4$ spinors for these four external legs take a simple form:
\eq
(\lambda_i)^{A}_{\;\;\;a}=\left(\begin{array}{cc}0 & \lambda_{i\alpha} \\ \tilde{\lambda}_i^{\dot{\alpha}} & 0\end{array}\right)\,,\quad (\tilde{\lambda}_i)_{A\dot{a}}=\left(\begin{array}{cc}0 & \lambda_{i}^\alpha \\- \tilde{\lambda}_{i\dot{\alpha}} & 0\end{array}\right)\,.
\eqe 
This requires us to split the six-dimensional momenta into a four-dimensional piece and two extra dimensions, $ (k^{(6)})^2=(k^{(4)})^2+m\tilde{m}$ with $m=k_4+ik_5,\;\; \tilde{m}=k_4-ik_5$. 
With this representation the loop-momentum of a double-cut can be parametrized as:
\beqa
\label{param}
\begin{array}{lll}
\displaystyle(\la_\ell)^{A}_{\;\;\; a}=\left(\begin{array}{cc}-\ka \la_j&\displaystyle\frac{(1-y)}{c} \la_j + 
     \la_{i}\\
   c \tilde \la_j + 
     y \tilde\la_{i}\,,& \displaystyle \frac{\tilde \ka}{c} \tilde\la_{i}\end{array}\right) \,,&\displaystyle\ka=\frac{m}{\langle ij\rangle}\,, &\displaystyle\tilde\ka=\frac{\tilde m}{[ij]}\,,\\
\displaystyle(\tilde\la_\ell)_{A\dot a} =\left(\begin{array}{cc}
\ka'\la_j& \displaystyle\frac{w(1-y)}{c} \la_j+\la_{i}\\
-c \tilde\la_{j}-y \tilde\la_{i}&\displaystyle\frac{\tilde \ka'}{c} \tilde \la_{i}
\end{array}\right)\,,&\displaystyle\ka'=\frac{\tilde m}{\langle ij\rangle}\,,&\displaystyle\tilde\ka'=\frac{m}{[ ij]} \,, 
\end{array}
\eeqa
where the position of the loop-momentum and the legs associated with $i,j$ are indicated in the bubble diagram of figure~\ref{2-cut}. To obtain the triple- and quadruple-cut integrands without using the three-point amplitude, we will instead use the four-point amplitude and multiply it by an inverse propagator. When the kinematics are such that the propagator becomes on-shell, one automatically obtains the product of 2 three-point amplitude. We will start with the double-cut and the extraction of the bubble coefficients, and by puting one and two more propagators on-shell, we extract the wanted triangle and box coefficients, respectively.

Using the parameterization in eq.\eqref{param}, the double-cut integrands are functions of $c$, $y$ $m$ and $\tilde{m}$. We will use the two parameters $y$ and $c$ to probe the structure of the propagators in the cut. Since the integrand is given by a product of tree-amplitudes, it is simply a rational function and one can express it as some Laurent series in both $c$ and $y$. There are four major contributions, (1) terms where there are no poles in both $c$ and $y$ (2) terms where there is poles in $c$ (3) terms where there are poles in $y$, and (4) terms where there is poles in both $y$ and $c$. Case (1) obviously contributes to the scalar bubble integral. For case (4), the presence of two poles simply represent two extra propagators, and correspond to box integrals, and thus do not contribute to the bubble coefficient. For case (2) and (3) the single pole indicates a triangle integral, however, due to the possible $y$ and $c$ dependence of the residue respectively, the can represent loop-momentum dependent numerators that can contribute to the scalar bubble integral via integral reduction. Without loss of generality, we will use the $y$-parameter to probe this triple-cut singularity. Thus the total contribution to the bubble integral can be obtained from 
\beqa
\left\{\int dcdy J_{c,y} [\mathrm{Inf}_c[\mathrm{Inf}_y A_1A_2](y)](c)\right\}+\left\{\frac{1}{2}\sum_{\{y=y_{\pm}\}}\int dc J'_{c}[\mathrm{Inf}_cA_1A_2A_3](c)\right\}\,, 
\eeqa
where $\mathrm{Inf}_x$  is an operator that expands its argument as a polynomial of $x$ at infinity, i.e. dropping the pole terms in the Laurent expansion. The Jacobian factor $J_{c,y}$ arises from the change of coordinates from the original loop momentum, subject to on-shell constraints, to eq.\eqref{param}. Similarly for $J'_{c}$, except that now there is one additional propagator that has been put on-shell, thus fixing $y=y_{\pm}$, the two triple-cut solutions. In the sum $\sum_{\{y=y_{\pm}\}}$, one is summing over  all possible scalar triangles which also share the same double-cut as scalar bubbles. The explicit form $J_{c,y}$ and $J'_c$ is given in~\cite{FordeRules}, and it can be shown that the first integral only picks up non-trivial contributions from terms of the form $c^0y^m$ where $m\geq0$, whilst the second integral one only consider terms of the form $c^n$ with $n>0$. However, different powers of $c$ correspond different powers of loop momenta which can be reduced to the integral basis of scalar bubbles and scalar triangles via Passarino-Veltman reduction. As the natural of specific parametrization of loop momenta, one simply replaces the integrations over $c^n$ with $\mathfrak{C}_n$ to obtain the contributions of bubble coefficients, where $\mathfrak{C}_n$ are the explicit results of integrating over different powers of the parameters $c$. They are given by:
\beqa\label{CCrap}
\label{Cn}\mathfrak{C}_n=\left(-\frac{\langle \chi|K_2|K_{1}^p]S_1}{\ga\De}\right)^n \mathcal{T}_n\,,
\eeqa
where
\beqa
\notag \mathcal T_1&=&-\frac{1}{2}\,,\\
\notag \mathcal T_2&=&-\frac{3 }{8 } \left[S_2+(K_1\cdot K_2)\right]\,,\\
\notag \mathcal T_3&=&- \frac{S_2}{48}\left[4 S_1+15 S_2+30(K_1\cdot K_2)+\frac{11
 (K_1\cdot K_2)^2}{S_2}+\frac{16 
m\tilde m \De}{S_1S_2}\right]\,,\\
\notag\mathcal T_4&=&-\frac{5S_2}{384}  \left[S_2+(K_1\cdot K_2)\right]\times \\
\notag &&~~~~~~~\left[ 11 S_1+21 S_2+42 
(K_1\cdot K_2)+\frac{10  (K_1\cdot K_2)^2}{S_2}+\frac{44 m\tilde m \De}{S_1S_2}\right]\,,\\
\notag \mathcal T_5&=&-\frac{S_2^2}{3840}
\bigg\{ \left(64 S_1^2+735 S_1 S_2+945 S_2^2\right)+210 
(7 S_1+18 S_2)(K_1\cdot K_2)\\
\notag &&+ \frac{(607 S_1+4935 S_2)(K_1\cdot K_2)^2}{S_2}+\frac{2310 (K_1\cdot K_2)^3}{S_2}+\frac{274(K_1\cdot K_2)^4}{S_2^2}\\
\notag &&+\frac{4 m\tilde m\Delta}{S_1 S_2 }  \left[ 128 S_1+735 S_2+1470 (K_1\cdot K_2)+\frac{607
(K_1\cdot K_2)^2}{S_2}\right]+\frac{1024 (m\tilde m)^2  \Delta ^2}{S_1^2S_2^2}\bigg\}\,,\\
\notag\mathcal T_6&=&\frac{S_2^2}{5120}\left[S_2 + (K_1\cdot K_2)\right]\bigg\{-7 \bigg[(33 S_1^2 + 170 S_1 S_2 + 165 S_2^2)\\
\notag&&+ 20  (17 S_1 + 33 S_2) (K_1\cdot K_2) +  \frac{ 4 (26 S_1 + 205 S_2) (K_1\cdot K_2)^2}{S_2}\\
\notag&& + 
       \frac{320  (K_1\cdot K_2)^3}{S_2} +\frac{ 28 (K_1\cdot K_2)^4)}{S_2^2}\bigg]  - \frac{
 3696 (m\tilde m)^2 \Delta^2}{S_1^2 S_2^2}\\
  \notag =     &&- 
 \frac{56  m\tilde m\Delta}{S_1S_2} \left[33 S_1 + 85 S_2 + 
    170 (K_1\cdot K_2) + \frac{52 (K_1\cdot K_2)^2}{S_2}\right]\bigg\}\,,\\
    &&\\
   \notag &&~~~~~~~~~~~~~S_i=K_i^2\,,~~\Delta=(K_1\cdot K_2)^2-S_1S_2\,,
\eeqa
and $\langle \chi|$ is a reference spinor that is used to define the null vector $K^p_1\equiv K_1-\chi S_1/(2K_1\cdot\chi)$. Here $K_1$ is the momenta of the original two-particle cut, and $K_2=k_q$ is the massless leg that has been separated to form a massless corner of the triple cut, as shown in fig.\ref{2-cut}. Thus the bubble coefficient is given by:
\beqa
\label{cbub}c_{bub}&=&[\mathrm{Inf}_c[\mathrm{Inf}_y A_1A_2](y)](c)\Big|_{c\rightarrow 0, y^m\rightarrow Y_m}+\frac{1}{2}\sum_{\{y=y_{\pm}\}}[\mathrm{Inf}_cA_1A_2A_3](c)\Big|_{c^n\rightarrow \mathfrak{C}_n}\,, 
\eeqa
where $Y_m$ are the results of integrating over different powers of the parameters $y$. They are given by:
\beqa
\notag &&~~~Y_0=1\,,~~~~~Y_1=\frac{1}{2}\,,~~~~~Y_2=\frac{1}{3}\left(1-\frac{m\tilde m}{S_1}\right)\,,\\\label{Ym}&&Y_3=\frac{1}{4}\left(1-\frac{2m\tilde m}{S_1}\right)\,,~~~~~~Y_4=\frac{1}{5}\left[1-\frac{3 m\tilde m}{S_1}+\frac{(m\tilde m)^2}{S_1^2}\right]\,.
\eeqa
Note that due to the lack of color-ordering, for a given bubble coefficient the number of triple-cuts that might contribute to the double-cut is four for QED and gravity, as opposed to two for QCD. Moreover choosing $\langle \chi|=\langle k_j|$, $\mathfrak{C}_n \propto (\langle k_j k_q\rangle[k_q k_i])^n$, and we have $\mathfrak{C}_n=0$ when ever the chosen massless corner $k_q$  is identified with $k_i$ or $k_j$. For QCD, amongst the two-triangles there is one for which such a configuration occurs, and thus only one remaining triangle integral contributes to the bubble coefficient. As for QED and gravity, amongst the four possible triangle configurations exactly two of them correspond to $k_q=k_i$ and $k_q=k_j$ respectively. Thus only two triangles contribute to the bubble coefficient. 
\par 

\begin{figure}[h]
\centering
\includegraphics[scale=0.7]{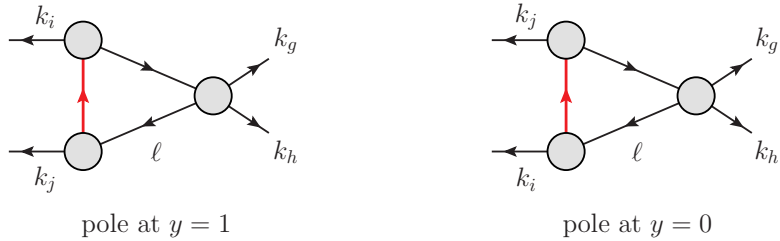}
\caption{The propagators marked with red lines in the left and right diagrams have poles at $y=1$ and $y=0$, respectively.}\label{3-cut}
\end{figure}

To extract the triangle coefficients, we set y to a fixed value such that an additional propagator goes on shell. 
As shown in fig.\ref{3-cut} $y=1$ or $0$ correspond to two distinct ordered triangles. After fixing $y$, we can expand the triple-cut integrand by $c$ around infinity and the triangle coefficients can be written as a polynomial of $c$ up to some powers. For the choice of in eq.\eqref{param}, it turns out only zero power of $c$ contributes. Note that for fixed $y$, we must average over loop momenta parametrized by $c$ and its complex conjugate  $c^*$. So the triangle coefficients are given by
\beqa
\label{ctri}c_{tri}=\frac{1}{2}\left([\mathrm{Inf}_cA_1A_2A_3](c)\Big|_{c=0}+[\mathrm{Inf}_cA_1A_2A_3](c*)\Big|_{c^*=0}\right)\,.
\eeqa
Note that, like in the situation of bubble coefficients, for theories without color ordering the two diagrams as shown in fig.\ref{3-cut} contribute to the same triangle coefficient. 
\par

Finally, having set $y=\tilde y$, where $\tilde y=0$ or $1$, we can imposing one more constraint to obtain the box coefficients. The one more constraint is a quadratic equation in $c$. We can solve this constraint to have $c=c_{\pm}(\tilde y)$. So the parameters $c$ and $y$ are completely fixed as well as the box coefficients
\beqa
c_{box}=\frac{1}{2}\left[(A_1A_2A_3A_4)_{c=c_+(\tilde y)}+(A_1A_2A_3A_4)_{c=c_-(\tilde y)}\right]\,,
\eeqa  
where we average over two solutions of $c$. Again for the non-color-ordered theories, there are two equal contributions to a given box coefficient, one is from substituting $c_{\pm}(y=0)$ and the other $c'_{\pm}(y=1)$, where $c$ and $c'$ are solutions for the on-shell condition of different propagators. These two configurations correspond to clockwise and counterclockwise flow of loop momenta.\par

Before finishing, we need to address how to deal with the $m\tilde m$ terms. The appearance of these terms is an artifact of our insistence  on putting the four-external legs on a manifest four-dimensional subspace. Since $m\tilde m$ is the extra component of the loop momentum that is orthogonal to this subspace, we can write $m\tilde m$ as 
\eq
m\tilde m=-16 \frac{Gram(\ell, k_1,k_2,k_3,k'_4)}{(stu)^2}\,,
\eqe
where $Gram$ is the determinant of the Gram matrix, $ k_1,k_2,k_3$ and $k'_4=\epsilon^{\mu\nu\rho\sigma}k_{1\mu}k_{2\nu}k_{3\rho}$ now completely spans this four-dimensional sub plane. Using reduction, one can then show that, for general $D>4$,
\beqa\label{mmCrap}
 I_{4m}[3]&=&
-\frac{D(D-2)(D-4)}{16(D^2-1)}
\left[
\frac{\mathcal{I}_{34}}{(D-3)}
-\sum_{ x=s,t}\frac{(2 u-2D u-D x) x^2 I_2[x]}{(D-2)D  u^2}
\right]\,,\\
 I_{4m}[2]&=&\frac{(D-2)(D-4) }{4
 (D-1)}\left[\frac{u}{s t}\frac{\mathcal{I}_{34}}{(D-3)}-\sum_{x=s,t}\frac{ x I_2[x]}{(D-2) u}\right]\,,\\
\label{4m1}I_{4m}[1]&=&-\frac{(D-4)}{(D-3)}\frac{u^2}{s^2 t^2}\mathcal{I}_{34}\\
\mathcal{I}_{34}&\equiv&\frac{s^3 t^3}{4u^3}\left( I_4[s,t]-\sum_{ x=s,t}^{i=1,2}\frac{x I_3[x_i] }{s t}\right)\,,\\
I_{3m}[2,x_i]&=&-\frac{(D-2)(D-4)}{8D(D-1)}x I_2[x]\,,\\
I_{3m}[1,x_i]&=&-\frac{(D-4)}{2(D-2)}I_2[x]\,,\\
I_{2m}[1,x]&=&\frac{(D-4)}{4(D-1)}x I_2[x]\,,
\eeqa
where
\beqa
  I_{4m}[n]&\equiv&\int d^D\ell \frac{(m\tilde m)^n}{\ell^2(\ell-k_2)^2(\ell-k_2-k_3)^2(\ell+k_1)^2} \,,\\
\label{i3m}I_{3m}[n,x]&=&\int d^D\ell \frac{(m\tilde m)^n}{\ell^2(\ell-K_1)^2(\ell+K_2)^2} \,,~~2 K_1\cdot K_2=-x\,,\\
 I_{2m}[n,y]&\equiv&\int d^D\ell \frac{(m\tilde m)^n}{\ell^2(\ell-K)^2}\,,~~K^2=y\,,
\eeqa
and $I_3[x_i]$, $i=1,2$, are the two distinct $x$-channel scalar triangles. 

In the following, we will compute the parity-odd one-loop integral coefficients for chiral-fermion loop as an example of how the above procedures are carried out. 

\section{Example: 6D chiral and anti-chiral fermions in the loop \label{UnitarityAppendix}}
Here we consider 6D chiral and anti-chiral fermions in the loop as an example with $k_i=k_1$ and $k_j=k_4$ in the paramerization eq.\eqref{param}. For $y=1$ the parametrization becomes a triple cut, where as $(y=1, c=c_{\pm})$ with $c_{\pm}=\frac{\langle 12\rangle}{2\langle 42\rangle }\left(1\pm \sqrt{1+\frac{4m\tilde m u}{st}}\right)$ are the two quad-cut solutions. All of this is illustrated in the following:
$$\includegraphics[scale=0.69]{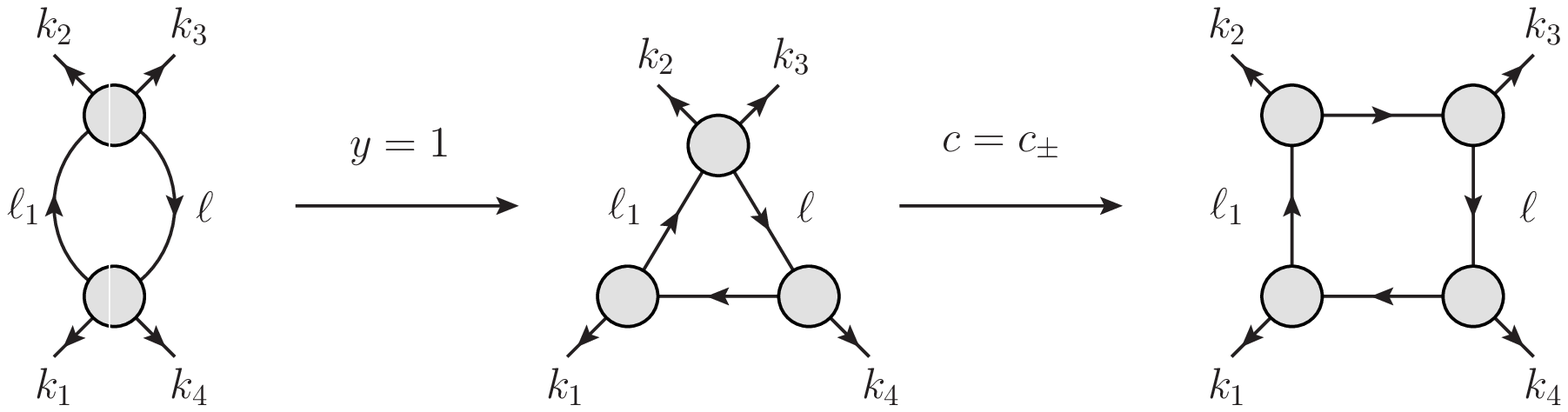}\,\,\,\,.$$
We will show that the cut-constructible parity-odd amplitude is given by scalar integrals with coefficient 
\eqa
\nonumber C_4&=&\frac{(s-t)}{6u^2}F^{(4)},\;\;C_{3s}=-\frac{(s-t)}{6tu^2}F^{(4)},\;\;C_{3t}=-\frac{(s-t)}{6su^2}F^{(4)}\\
C_{2s}&=&\frac{1}{stu}F^{(4)},\;\;C_{2t}=-\frac{1}{stu}F^{(4)}\,,
\label{CoefficientSol}
\eqae
where the coefficients for the triangles are identical for $I_3(1|2|34)$ and $I_{3}(12|3|4)$ and $F^{(4)}$ is given in eq.(\ref{F4Def}).

The computation is straightforward: we compute the unitarity cut result for both chiral and anti-chiral fermions in six-dimensions, and show that the difference of the two, which would be parity odd gives precisely the integral coefficients in the above. As the six-dimensional gluon becomes the four-dimensional gluon and two scalars, we will choose the $A(\phi_1,\bar{\phi}_2,3^-,4^+)$ component amplitude. In 6D notation this correspond to $A(1_{1\dot{2}}\,,2_{2\dot{1}}\,,3_{2\dot{2}}\,,4_{1\dot{1}})$, and for later reference, we have 
\eq
F^{(4)}(1_{1\dot{2}}\,,2_{2\dot{1}}\,,3_{2\dot{2}}\,,4_{1\dot{1}})=s\langle 23\rangle^2[24]^2\,.
\label{F4Id}
\eqe

\subsection{Chiral fermion and anti-chiral fermion loop}
First we consider the double-cut, the cut integrand is
\eq
\label{ex2cut}A((-\ell_1)_a 2_{2\dot{1}}3_{2\dot{2}}\ell_b)A((-\ell)^b4_{1\dot{1}}1_{1\dot{2}}\ell_1^a)\,,
\eqe
where the tree-amplitude for a chiral fermion is given by 
\eqa
\nonumber A((-\ell_1)_a2_{2\dot{1}}3_{2\dot{2}}\ell_b)&=&\frac{\langle (-\ell_1)_a2_{2}3_{2}\ell_b\rangle[\ell_{\dot{b}}2_{\dot{1}}3_{\dot{2}}\ell^{\dot{b}}]}{ts_{-\ell_12}}\,,\\
A((-\ell)_b4_{1\dot{1}}1_{1\dot{2}}(\ell_1)_a)&=&\frac{\langle (-\ell)_b4_{1}1_{1}(\ell_1)_a\rangle[\ell_{\dot{b}}4_{\dot{1}}1_{\dot{2}}\ell^{\dot{b}}]}{ts_{\ell_11}}\,.
\eqae
For the bubble coefficient, consider eq.\eqref{cbub}. Direct substitution of the loop momentum parameterization into the above tree-amplitudes and expand in $y$ and $c$,  one finds that the first term in eq.\eqref{cbub} receives no contributions. For the second term one finds a polynomial in linear in $c$ and thus will give non-vanishing contribution. Using  eq.(\ref{CCrap})
\beqa
\mathfrak{C}_1=\frac{2\langle13\rangle[34]}{t^2}\,,
\eeqa
one finds the following bubble coefficient:
\beqa
C_{2t}^-=\frac{(2 m\tilde m-t)}{ t^2 u}\langle 2, 3\rangle^2[2,4]^2\,.
\eeqa
where the $^-$ refers to the chiral contribution. To obtain the triangle coefficient, first we have triple-cut integrand by simply multiplying eq.(\ref{ex2cut}) with the propagator $(\ell_2-k_1)^2$ and set the on-shell condition $y=1$
\beqa
\label{ex3cut}A((-\ell_1)_a 2_{2\dot{1}}3_{2\dot{2}}\ell_b)A((-\ell)^b4_{1\dot{1}}1_{1\dot{2}}\ell_1^a)(\ell_1+k_1)^2\big|_{y=1}\,.
\eeqa
The triangle coefficient can be obtained by taking $c\rightarrow\infty$ for eq.\eqref{ex3cut} to extract the $\mathcal{O}(c^{0})$ piece. We will also need to do this for the conjugate momenta, but there is no pole at infinity for the conjugate solution. In the end the triangle coefficient is given as
\eqa &&C_{3t}
=-\frac{\langle 23\rangle^2[24]^2 t}{2u^2}\left(\frac{2m\tilde{m}}{t}-1\right)\,.
\eqae
Similarly, the quad-cut integrand is given by
\eq
A((-\ell_1)_a 2_{2\dot{1}}3_{2\dot{2}}\ell_b)A((-\ell)^b4_{1\dot{1}}1_{1\dot{2}}\ell_1^a)(\ell_1-k_2)^2(\ell_1+k_1)^2\big|_{y=1,c=c^{\pm}}\,.
\eqe
Finally the box coefficient is obtained by averaging over $c=c_+$ and $c=c_-$,
\eq
C_{4}^-=\left(\frac{2m\tilde{m}}{t}-1\right)\langle23\rangle^2[24]^2\left(\frac{m\tilde{m}}{u}+\frac{st}{2u^2}\right)\,.
\eqe
To get the parity odd term, we want to subtract this against the anti-chiral fermion loop cut, which is given by 
\eq
A((-\ell_1)_{\dot{a}} 2_{2\dot{1}}3_{2\dot{2}}\ell_{\dot{b}})A((-\ell)^{\dot{b}}4_{1\dot{1}}1_{1\dot{2}}\ell_1^{\dot{a}})\,.
\eqe
Following similar steps we obtain
\beqa
C_{2t}^+&=&\frac{2 m\tilde m}{t^2 u}\langle 1, 3\rangle^2[1,4]^2\,,\\
C_{3t}^+&=&-\frac{\langle 23\rangle^2[24]^2}{u^2}m\tilde{m}\,,\\
C_{4}^+&=&\left(\frac{m\tilde{m}}{t}\right)\langle23\rangle^2[24]^2(-1+\frac{2m\tilde{m}}{u}+\frac{st}{u^2})\,.
\eeqa

\subsection{The final result}
We now take the difference of the two cuts. For the quad-cut, one has 
\eq
C_{4}^--C_{4}^+=\langle 23\rangle^2[24]^2\left[\left(\frac{1}{t}-\frac{1}{u}\right)m\tilde m-\frac{st}{2u^2}\right],\,
\label{quadcutres}
\eqe 
where as 
\eq
C_{3t}^--C_{3t}^+=\frac{\langle 23\rangle^2[24]^2 t}{2u^2} 
\eqe 
and 
\eq
C_{2t}^--C_{2t}^+=-\frac{\langle 23\rangle^2[24]^2 }{t u}\,.
\eqe
Thus plugging eq.\eqref{4m1} into eq.(\ref{quadcutres}), we find the box coefficient and the $t$-channel triangle and bubble coefficients are given as:
\eq
C_4=\frac{s(s-t)}{6u^2}\langle 23\rangle^2[24]^2\,, \quad  C_{3t}=\frac{(t-s)}{6u^2}\langle 23\rangle^2[24]^2\,,\quad C_{2t}=-\frac{1}{tu}\langle 23\rangle^2[24]^2\,,
\eqe
which matches eq.(\ref{CoefficientSol}) with the identification of $F^4$ in eq.(\ref{F4Id}).
One can also obtain the bubbles by using the fact that the answer must be UV finite and read of the coefficient from the triangles.


\section{Results of integral coefficients}\label{RIC}
Here we list all the remaining integral coefficients for chiral-gravitino and chiral-fermion. We first list the $m\tilde{m}$-dependent integral coefficients. While they are only an intermediate step in getting the final integrals, they can be useful in obtaining four-dimensional rational terms as discussed in~\cite{YT6D, Scott6D}. The $m\tilde m$-dependent integral coefficients for chiral-gravitino are
\beqa
\notag &&\tilde C_4^\psi(1,2,3,4) = \frac{R^{(4)}}{t^2 u^4} [10  u^2 (t-u) (m\tilde m)^3-
     s u(4 s^2 + 7 s t - 17 t^2)(m\tilde m)^2    \\
\notag    &&~~~~~~~~~~~~~~~~~~~~~~~~~~~~~~~~   -t (2 s^4 + 10 s^3 t + 5 s^2 t^2 + 6 s t^3 + 4 t^4)m\tilde m +s t^3 (s^2 + t^2)]\,,
 \\
\notag &&\tilde C_4^\psi(1,3,4,2) = \frac{R^{(4)}}{ t^4 u^2}[ 10  t^2 (t- u)(m\tilde m)^3+ s t (4 s^2 + 7 s u - 17 u^2)(m\tilde m)^2\\
\notag&&~~~~~~~~~~~~~~~~~~~~~~~~~~~~~~~~ +
     u (2 s^4 + 10 s^3 u + 5 s^2 u^2 + 6 s u^3 + 4 u^4)m\tilde m-s u^3 (s^2 + u^2) ]\,,\\
\notag &&\tilde C_4^\psi(1,3,2,4) =\frac{2 (t - u) m\tilde m[5 (m\tilde m)^2 + 2 s  m\tilde m+ t u ]R^{(4)}}{t^2 u^2}\,,\\
\notag &&\tilde C_{3s}^\psi =\frac{2(t - u)R^{(4)}}{t^4 u^4} [-
     (5 t^5 + 19 t^4 u + 28 t^3 u^2 + 28 t^2 u^3 + 19 t u^4 +      5 u^5)m\tilde m \\
\notag&&~~~~~~~~~~~+ 
    s^2 (2 t^4 + 2 t^3 u + 3 t^2 u^2 + 2 t u^3 + 2 u^4) +5 t u (2 t^2 + 3 t u + 2 u^2) (m\tilde m)^2]\,,\\
\notag&&\tilde C_{3t}^\psi = -\frac{ 2[ 
  5 u(2 t-u)(m\tilde m)^2+ t^2 (5 s - 4 u)m\tilde m + t^2 (s^2 + t^2)]R^{(4)}}{ t u^4}\,,\\
\notag&&\tilde C_{3u}^\psi =-\frac{2[ 
   5 t(t-2 u) (m\tilde m)^2 - u^2 (5 s - 4 t)m\tilde m - u^2 (s^2 + u^2)]R^{(4)}}{ t^4 u}\,,\\
\notag&&\tilde C_{2s}^\psi =\frac{(t - u) [ 
   5 (2 t^2 + 3 t u + 2 u^2) m\tilde m+2 s (2 t^2 + t u + 2 u^2)]R^{(4)}}{t^3 u^3}\,,\\
\notag&&\tilde C_{2t}^\psi = \frac{(s- t)(5 m\tilde m - 2 t)R^{(4)}  }{t u^3}\,,\\
&&\tilde C_{2u}^\psi =-\frac{(s- u)(5 m\tilde m - 2 u)R^{(4)} }{t^3 u} \,.
\eeqa
and for the chiral fermion are
\beqa
\notag &&\tilde C_4^\chi(1,2,3,4) = \frac{   [2  u^2(t - u) (m\tilde m)^3+s t u(4 t - u)(m\tilde m)^2  +s^2 t^3 m\tilde m]R^{(4)} }{
 t^2 u^4}\,,
 \\
\notag &&\tilde C_4^\chi(1,3,4,2) =\frac{ [2  t^2 (t - u)(m\tilde m)^3 +  s tu (t - 4 u) (m\tilde m)^2 - s^2 u^3 m\tilde m]R^{(4)}}{ t^4 u^2}\,, \\
\notag &&\tilde C_4^\chi(1,3,2,4) =\frac{ 2(t - u)(m\tilde m)^3R^{(4)}}{t^2 u^2}\,,\\
\notag &&\tilde C_{3s}^\chi =\frac{2(t - u)  [ t u (2 t^2 + 3 t u + 2 u^2)(m\tilde m)^2 + s^3 (t^2 + u^2)m\tilde m]R^{(4)}}{t^4 u^4}\,,\\
\notag&&\tilde C_{3t}^\chi =-\frac{2[ u(2 t - u)  (m\tilde m)^2+ s t^2m\tilde m]R^{(4)}  }{t u^4}\,, \\
\notag&&\tilde C_{3u}^\chi =-\frac{2[ t (t - 2 u)(m\tilde m)^2 - s u^2 m\tilde m]R^{(4)}}{ t^4 u}\,,\\
\notag&&\tilde C_{2s}^\chi =\frac{ (2 t^3 + t^2 u - t u^2 - 2 u^3)m\tilde m R^{(4)}}{t^3 u^3}\,,\\
\notag&&\tilde C_{2t}^\chi = \frac{( s -t)m\tilde m R^{(4)}}{t u^3}\,,\\
&&\tilde C_{2u}^\chi =- \frac{ (s -u)m\tilde mR^{(4)}}{t^3 u}\,.
\eeqa
After substituting the results in eq.(\ref{mmCrap}) for distinct powers of $m\tilde m$, the parity-odd integral coefficients of chiral-gravitino loops are
  \beqa
\notag &&C_4^{\psi}(1,2,3,4)=\frac{(25 t^5 + t^4 u - 63 t^3 u^2 - 85 t^2 u^3 - 67 t u^4 - 
  21 u^5)R^{(4)}}{105 u^5}\,,\\
\notag &&C_4^{\psi}(1,3,4,2)=\frac{(21 t^5 + 67 t^4 u + 85 t^3 u^2 + 63 t^2 u^3 - t u^4 - 25 u^5)R^{(4)}}{105 t^5}\,,\\
 \notag &&C_4^{\psi}(1,4,2,3)=-\frac{(14 t^3 + 19 t^2 u - 19 t u^2 - 14 u^3)R^{(4)}}{70 s^3}\,,\\
 \notag &&C_{3s}^{\psi}=-\frac{2 s^2 (25 t^7 - 49 t^6 u + 10 t^5 u^2 - 56 t^4 u^3 + 56 t^3 u^4 - 
   10 t^2 u^5 + 49 t u^6 - 25 u^7)R^{(4)}}{105 t^5 u^5}\,,\\
 \notag  &&C_{3t}^{\psi}=-\frac{t (50 t^6 + 102 t^5 u - 72 t^4 u^2 - 420 t^3 u^3 - 642 t^2 u^4 - 
    537 t u^5 - 161 u^6)R^{(4)}}{105 s^3 u^5}\,,\\
 \notag  &&C_{3u}^{\psi}=-\frac{u (161 t^6 + 537 t^5 u + 642 t^4 u^2 + 420 t^3 u^3 + 72 t^2 u^4 - 
   102 t u^5 - 50 u^6)R^{(4)}}{105 s^3 t^5}\,,\\
 \notag &&C_{2s}^{\psi}=\frac{(150 t^6 - 244 t^5 u - 63 t^4 u^2 + 63 t^2 u^4 + 244 t u^5 - 
   150 u^6)R^{(4)}}{210 t^4 u^4}\,,\\
  \notag &&C_{2t}^{\psi}=-\frac{(150 t^4 + 106 t^3 u - 399 t^2 u^2 - 501 t u^3 - 56 u^4)R^{(4)}}{210 s^2 u^4}\,,\\
 &&C_{2u}^{\psi}=-\frac{(56 t^4 + 501 t^3 u + 399 t^2 u^2 - 106 t u^3 - 150 u^4)R^{(4)}}{210 s^2 t^4} \,. 
 \eeqa
  and the parity-odd integral coefficients of chiral-fermion loops are
\beqa
\notag &&C_4^{\chi}(1,2,3,4)=-\frac{s^3 t (5 t + 2 u)R^{(4)}}{105 u^5}\,,\\
\notag &&C_4^{\chi}(1,3,4,2)=\frac{s^3 u (2 t + 5 u)R^{(4)}}{105 t^5}\,,\\
\notag &&C_4^{\chi}(1,4,2,3)=-\frac{t u (t - u)R^{(4)} }{70 s^3}\,,\\
\notag &&C_{3s}^{\chi}=-\frac{2 s^4 (5 t^5 - 3 t^4 u + 3 t^3 u^2 - 3 t^2 u^3 + 3 t u^4 - 
   5 u^5)R^{(4)} }{105 t^5 u^5}\,,\\
\notag   &&C_{3t}^{\chi}=-\frac{t (10 t^6 + 54 t^5 u + 120 t^4 u^2 + 140 t^3 u^3 + 90 t^2 u^4 + 
   27 t u^5 + 7 u^6)R^{(4)} }{105 s^3 u^5}\,,\\
\notag   &&C_{3u}^{\chi}=\frac{u (7 t^6 + 27 t^5 u + 90 t^4 u^2 + 140 t^3 u^3 + 120 t^2 u^4 + 
    54 t u^5 + 10 u^6)R^{(4)} }{105 s^3 t^5}\,,\\
\notag &&C_{2s}^{\chi}=\frac{(30 t^6 + 52 t^5 u + 21 t^4 u^2 - 21 t^2 u^4 - 52 t u^5 - 
  30 u^6)R^{(4)}}{210 t^4 u^4}\,,\\
\notag   &&C_{2t}^{\chi}=-\frac{(30 t^4 + 122 t^3 u + 189 t^2 u^2 + 135 t u^3 + 56 u^4)R^{(4)}}{ 210 s^2 u^4}\,,\\
 &&C_{2u}^{\chi}=   \frac{(56 t^4 + 135 t^3 u + 189 t^2 u^2 + 122 t u^3 + 30 u^4)R^{(4)}} {
 210 s^2 t^4}\,.
 \eeqa


\end{document}